\documentclass[9pt,twocolumn,twoside]{osajnl}
\usepackage{caption}
\usepackage{amsmath}
\usepackage{csquotes}
\usepackage{hyperref}
\journal{ao} % Choose journal (ao, josaa, josab)

\setboolean{shortarticle}{false} % true = letter, false = research article

\title{Radiative transfer model for contaminated rough slabs}

\author[1,2,*]{Andrieu Fran\c{c}ois}
\author[3]{Dout\'{e} Sylvain}
\author[1,2]{Schmidt Fr\'{e}d\'{e}ric}
\author[3]{Schmitt Bernard}

\affil[1]{Universit\'{e} Paris-Sud, Laboratoire GEOPS, UMR8148, Orsay F-91405, France}
\affil[2]{CNRS, Orsay F-91405, France}
\affil[3]{Institut de Plan\'{e}tologie et d'Astrophysique de Grenoble, Grenoble 38041, France}

\affil[*]{Corresponding author: francois.andrieu@u-psud.fr}

\dates{Compiled \today}

\ociscodes{(010.5620) Radiative transfer ;  (240.6490) Spectroscopy, surface; (110.4234) Multispectral and hyperspectral imaging ; (240.5770) Roughness ; (290.7050)   Turbid media ; (080.2468)   First-order optics}

\doi{\url{http://dx.doi.org/10.1364/ao.XX.XXXXXX}}

\begin{abstract}
We present a semi-analytical model to simulate bidirectional reflectance
distribution function (BRDF) spectra of a rough slab layer containing
impurities. This model has been optimized for fast computation in
order to analyze hyperspectral data. We designed it for planetary
surfaces ices studies but it could be used for other purposes. It
estimates the bidirectional reflectance of a rough slab of material
containing inclusions, overlaying an optically thick media (semi-infinite
media or stratified media, for instance granular material). The inclusions
are supposed to be close to spherical, and of any type of other material
than the ice matrix. It can be any type of other ice, mineral or even
bubbles, defined by their optical constants. We suppose a low roughness
and we consider the geometrical optics conditions. This model is thus
applicable for inclusions larger than the considered wavelength. The
scattering on the inclusions is assumed to be isotropic. This model
has a fast computation implementation and thus is suitable for high
resolution hyperspectral data analysis.
\end{abstract}

\setboolean{displaycopyright}{true}

\begin{document}

\maketitle
\thispagestyle{fancy}
\ifthenelse{\boolean{shortarticle}}{\abscontent}{}

\section{Introduction}

Hyperspectral imaging has become a major component in planetary surface
observation since the past decades. Earth and other Solar system bodies
are now observed in various spectral ranges at various resolutions
and from various heights. 

As a photon come across a surface, it interacts in two major ways.
It can be either absorbed or deviated (scattering, diffraction, refraction).
The objective of this radiative transfer model is to describe the
interactions using a realistic surface description. In such descriptions,
the reflectance of a surface is the result of multiple interactions,
with multiple irregular interfaces of different materials. The exact
resolution of the radiative transfer equations turns out to be a highly
difficult and time consuming problem. This problem has been solved
under certain hypothesis : if the characteristic size of the particle
is much smaller than the wavelength, or if it is much bigger. In this
study, we consider the geometrical optics domain, where the particles
are much bigger than the wavelength. For example, in the visible and
near infrared range ($400\,\mbox{nm}-5\,\mbox{\textmu m}$), we suppose
that the average particle size does not fall below $10\,\mbox{\textmu m}$.
This is in general valid for planetary surfaces \cite{Peltoniemi1993}.
Ray tracing algorithms \cite{Grynko2003,Chang2005,Pilorget2013,Ben2014}
that simulate the very complex paths of millions of photons through
these surface can show very accurate results, but they depend on a
huge number of parameters and are highly time consuming, seriously
limiting the the interpretation of extensive hyperspectral images.
We aim at a radiative transfer model that is fast enough to be able
to deal with a vast amount of data, such as planetary spectro-imaging
databases. It is then necessary to make further simplifying assumptions
that enable the formulation of approximate analytic or semi-analytic
solutions to the radiative transfer problem. 
A possible simplification is to consider that the radiative properties
inside a media can be described statistically only using local mean
properties of scattering and absorption \cite{KUBELKA1948,Hapke1981,Shkuratov1999}.
The media is assumed to be homogenous at a mesoscopic scale. Another
classical simplifying assumption considered in such problem is the
two stream approximation \cite{KUBELKA1948,Hapke1981,Kylling1995}.
It has been shown that under certain conditions, it did not affect
too much the solution compared to more accurate studies, but simplifies
greatly the calculations \cite{Vargas1997,Vargas1999}. To describe
the reflectance of a surface, one also have to consider the geometry
of illumination and observation. In our approach, these photometric
effects are modeled by the properties of the interface between the
media and the exterior. These properties of roughness can also be
statistically described, using only one or a few parameters. Y. Shkuratov
\cite{Shkuratov1999} and B. Hapke \cite{Hapke1981} developed analytical
radiative transfer models for granular media, that are able to simulate
the bidirectional reflectance of various granular surfaces. 

If the media cannot be described as homogenous, it is possible to
consider it pieciwise continuus, constituted of homogeneous strata.
It is the case for example in the atmospheres, or stratified surfaces.
A family of models describe the radiative transfer in stratified media,
such as the DISORT algorithm \cite{Stamnes1988}. In these discrete-ordinate
modelisations, each layer is considered homogenous, and the total
reflectance is calculated iteratively, by adding the contribution
of each layer. This method has also been adapted to the study of the ocean-atmosphere coupled system with a rough surface  \cite{Jin2006}.

Starting from Hapke model, improving it, and combining it with a multi-layer
method \cite{Stamnes1988}, S. Dout\'{e} has also developed a model for
stratified granular surfaces \cite{Doute1998}. Using the same strategy,
we developed a semi-analytical radiative transfer model for a compact
layer (solid matrix containing inclusions) overlaying an optically
thick granular layer. This two layers approach does not require an
iterative DISORT-like method, but only adding coupling formulas. It
is founded on three major assumptions : (i) the geometric optics conditions
are observed, (ii) the media is piecewise continuous and (iii) the
inclusions are close to spherical and homogeneously distributed in
the matrix.

\subsection*{Model overview}

\begin{figure}[htbp]
\centering
\begin{center}
\fbox{(a)\includegraphics[width=\linewidth]{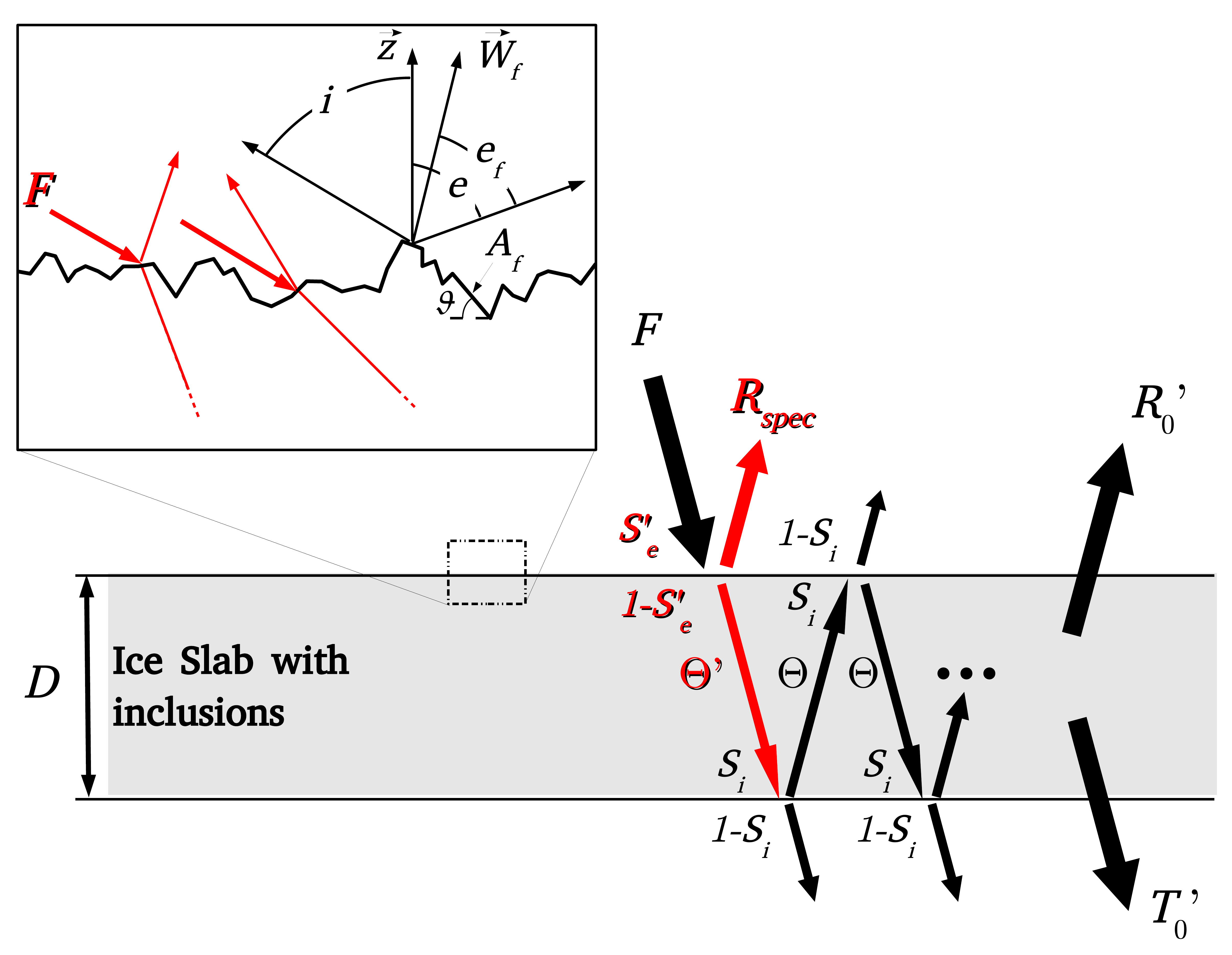}}
\end{center}
\begin{center}
\fbox{(b)\includegraphics[width=\linewidth]{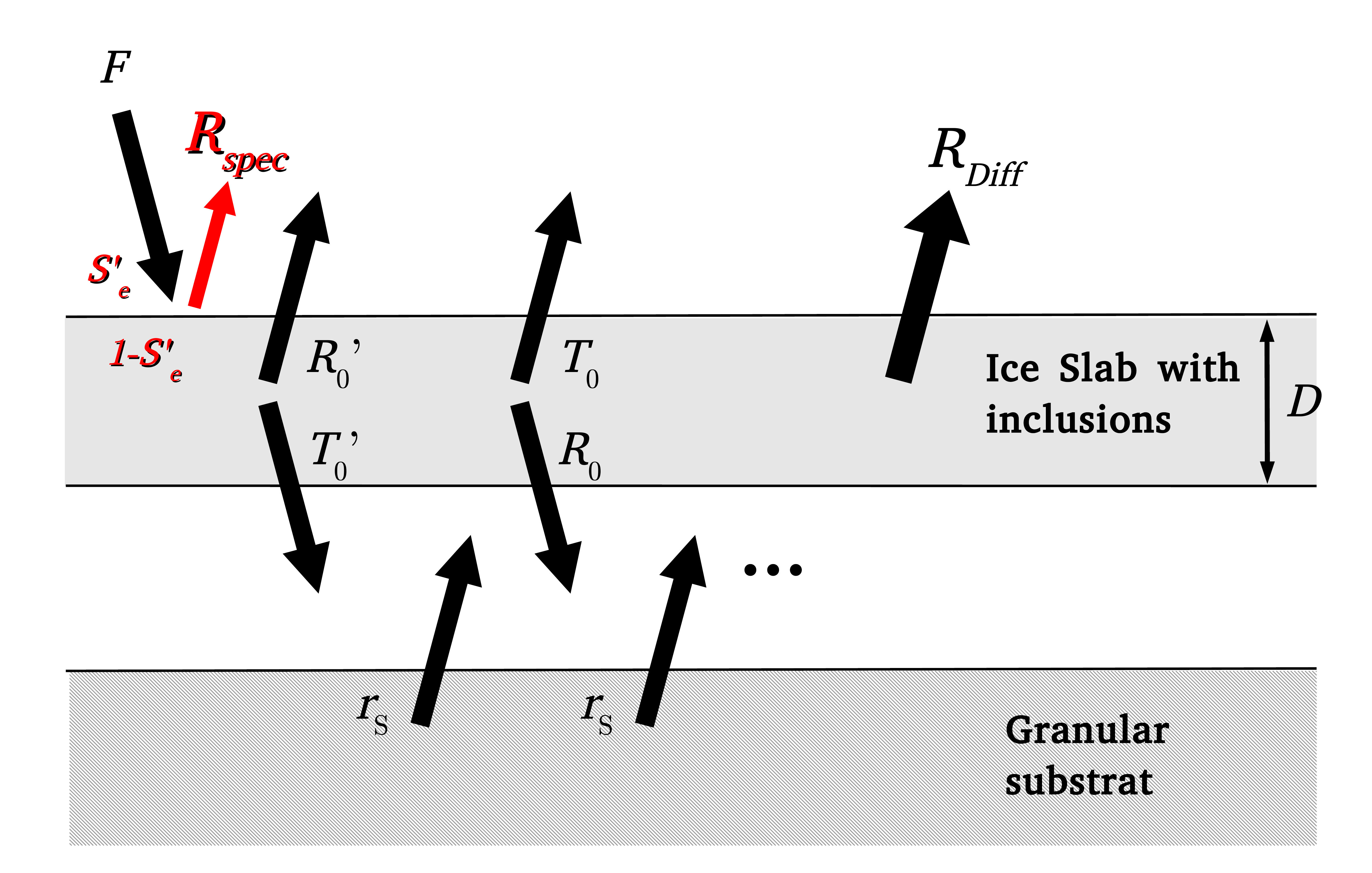}}
\end{center}

\caption{Illustration of the radiative transfer in a rough slab. (a) Radiative transfer for a slab ice layer only. Anisotropic transit are represented in red, and named with a prime. On the top left : illustration of the reflections and transmission at the first interface, used in the calculations of variables $S_{e}^{\prime}$ and $\overline{\mbox{fact}}$, $\Theta^{\prime}$. (b) Illustration of the adding coupling. The granular and slab layers are artificially separated in this figure to help the understanding of the coupling. }
\label{fig:Illustration-of-radiative}
\end{figure}

We decompose the reflectance into two distinct contributions : specular
and diffuse. We chose the Hapke \cite{Hapke1984} probability density
function of orientations, as it well describes the statistic distribution
of slopes in the approximation of small angles. We consider a collimated
incident radiation, at an incident angle $i$. We estimate the specular
contribution, considering the geometry and the surface description.
The specular reflection of rough surfaces have been studied in various
cases \cite{COX:54,Muhleman64,Saunders67,Hapke1984,vanGinneken:98}.
We use the same general idea of these methods, describing the rough
surface as constituted of multiple unresolved facets. The specular
contribution will result from the integration of the specular reflections
on the facets, in the solid angles considered (\textit{i.e.}, the
light source and the detector) as described in figure \ref{fig:Illustration-of-radiative}a.

Then we estimate the diffuse contribution. The total reflection coefficient
at the first rough interface, that determines the amount of energy
transmitted to the slab, is obtained by integrating specular contributions
in every emergent direction, at a given incidence. We consider that
the first transit through the slab is anisotropic (collimated), and
that there is an isotropisation at the second rough interface (\textit{i.e.
}when the radiation reach the semi-infinite substrate). For the refraction
and the internal reflection, every following transit is considered
isotropic. The diffuse contribution is obtained using an analytical
estimation of Fresnel coefficients \cite{Chandrasekhar1960,Doute1998},
and a simple statistical approach. The contribution of the semi infinite
substrate is estimated using Hapke model \cite{Hapkebook}. Finally,
we consider that the slab is under a collimated radiation from the
light source, and under a diffuse radiation from the granular substrate.
We compute the resulting total bidirectional reflectance using adding
doubling formulas (figure \ref{fig:Illustration-of-radiative}b).

\section{Surface roughness - Facets distribution}
 \label{sub:Surface-rugosity}
The first step is to describe the roughness of the surface. We consider
that it is composed of $N$ facets that are not resolved, with $N\gg1$.
These facets' orientations follow a probability density $\mbox{a}\left(\vartheta,\zeta\right)$,
where $\vartheta$ is the zenital angle between the normal to the
facet and the local vertical direction, and $\zeta$ is the azimutal
angle. To make our approach as general as possible, we chose to describe
the surface as randomly rough. Such a roughness has already been widely
studied (see for example \cite{COX:54,Muhleman64,Saunders67,Lumme81,Hapke1984,vanGinneken:98}
and the reference cited in these papers). These studies show that
a slope distribution, with $\tan\vartheta$ that is close to Gaussian
is a good description of the surface. Such a description combines
simplicity and efficiency reproducing the photometric variations.
For the sake of simplicity and because it is widely used in the literature,
we chose the following probability distribution function \cite{Hapke1984}
:

\begin{equation}
\mbox{a}\left(\vartheta,\zeta\right)=\frac{1}{\pi^{2}\tan^{2}\overline{\theta}}\exp\left(-\frac{\tan^{2}\vartheta}{\pi\tan^{2}\overline{\theta}}\right)\sec^{2}\vartheta\sin\vartheta\label{eq:a-dens-prob}
\end{equation}
where

\begin{equation}
\tan\overline{\theta}=\frac{2}{\pi}\int_{0}^{\frac{\pi}{2}}\mbox{a}\left(\vartheta\right)\tan\vartheta\,\mbox{d}\vartheta\label{eq:thetabar}
\end{equation}
It is supposed that the azimutal distribution is uniform. The angle
$\overline{\theta}$ representing the mean slope angle completely
characterizes the facets' orientations and the surface roughness.
This slopes distribution considers the cases of small $\bar{\theta}$.
Practically, the threshold of validity can be determined depending
on the level of tolerance (see Figures~\ref{fig:Error-in-cons} and
\ref{fig:Proba_slope}. The expression of $\mbox{a}\left(\vartheta,\zeta\right)$
could be adapted in the future to extend the study to any type of
terrain, as discussed in section~\ref{sub:Energy-conservation}. 

\begin{figure}[htbp]
\centering
\fbox{\includegraphics[width=\linewidth]{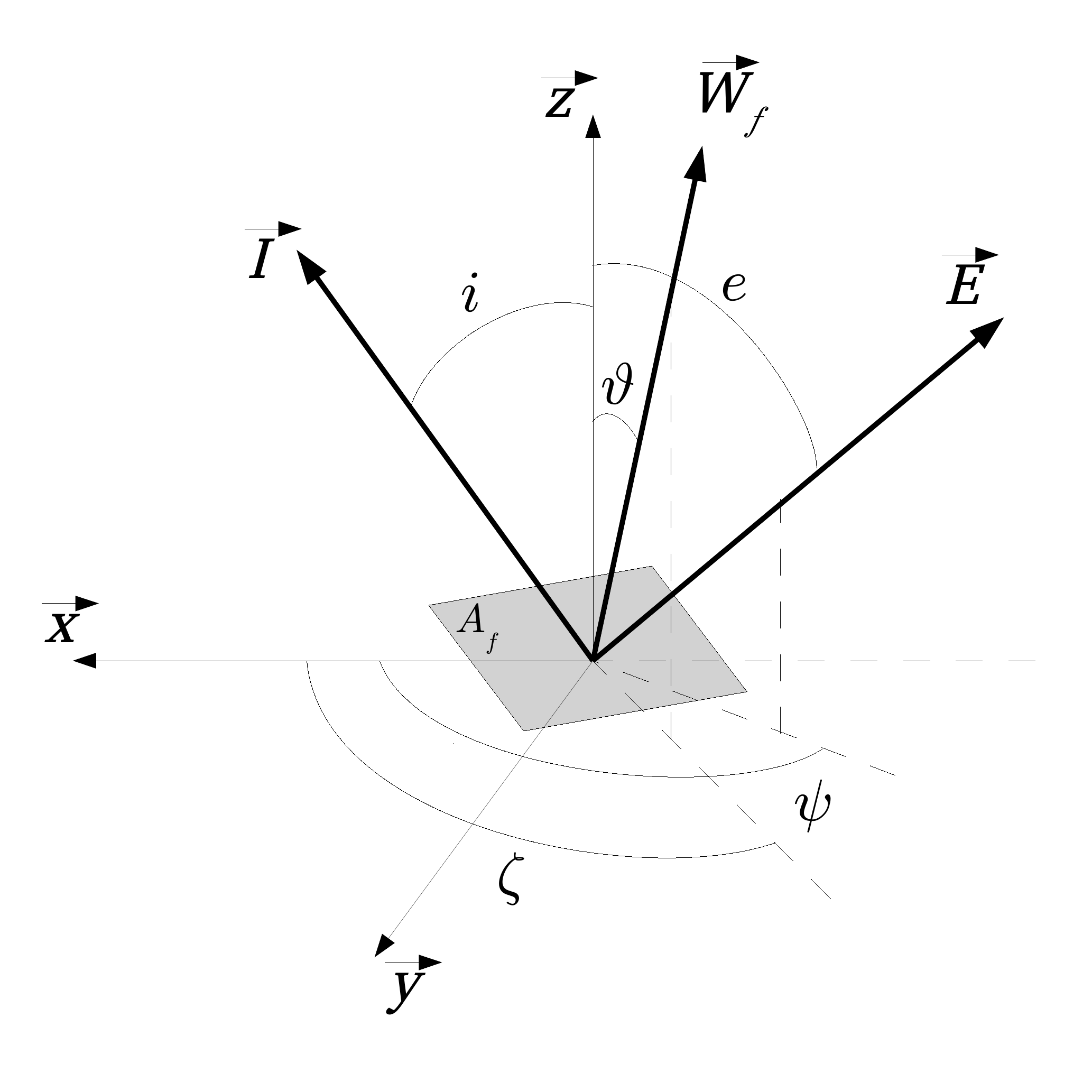}}
\caption{The local coordinate system $\left(\vec{x},\vec{y},\vec{z}\right)$
is centered at the slab surface, with the $\vec{z}$ axis vertically
upward and the $\vec{x}$ axis horizontally toward the sun. $\vec{I}$
and $\vec{E}$ are respectively the incident and the emergent directions.
$\vec{W_{f}}$ is the normal to the facet $A_{f}$.}
\label{fig:reperes}
\end{figure}

\section{Specular reflectance }

In this part we determine the specular contribution to the reflectance
at the detector. We first establish the relations between the orientation
$\left(\vartheta_{spec},\zeta_{spec}\right)$ of a facet that is in
specular conditions and the geometry of observation $\left(i,e,\psi\right)$,
where $i$ is the incidence angle, $e$ the emergence and $\psi$
the azimuth (see Figure~\ref{fig:reperes}). The total specular contribution
is obtained by integrating these relations, taking into account the
geometry variations within one pixel and the statistics of the slopes
$\mbox{a}\left(\vartheta,\zeta\right)$.

\subsection{Specular conditions for one facet}
\label{sub:Specular-conditions-for}

For one facet $A_{f}$ to satisfy the specular reflection conditions,
then its normal $\vec{W_{f}}$ must respect (see figure \ref{fig:reperes}):
\begin{equation}
\vec{E}=-\vec{I}+2\langle\vec{I}|\vec{W_{f}}\rangle\cdot\vec{W_{f}}\label{eq:spec}
\end{equation}
where the operator $\langle|\rangle$ represents the scalar product.
If we express Eq. \ref{eq:spec} in the $\left(\vec{x},\vec{y},\vec{z}\right)$
coordinates, : 
\begingroup\makeatletter\def\f@size{8}\check@mathfonts
\def\maketag@@@#1{\hbox{\m@th\large\normalfont#1}}
\begin{equation}
\begin{cases}
x_{e} =2\left(\sin i\sin\vartheta{}_{spec}\cos\zeta{}_{spec}+\cos\vartheta{}_{spec}\cos i\right)\sin\vartheta{}_{spec}\cos\zeta{}_{spec}-\sin i\\
y_{e} =2\left(\sin i\sin\vartheta{}_{spec}\cos\zeta{}_{spec}+\cos\vartheta{}_{spec}\cos i\right)\sin\vartheta{}_{spec}\sin\zeta{}_{spec}\\
z_{e} =2\left(\sin i\sin\vartheta{}_{spec}\cos\zeta{}_{spec}+\cos\vartheta{}_{spec}\cos i\right)\cos\vartheta{}_{spec}-\cos i
\end{cases}\label{eq:sys}
\end{equation}
\endgroup
where $x_{e}=\sin e\cos\psi $, $y_{e}=\sin e\sin\psi$ and $z_{e}=\cos e$ are the coordinates of the emergent
vector $\vec{E}$ in the $\left(\vec{x},\vec{y},\vec{z}\right)$ frame.
This leads to : 
\begin{equation}
\begin{cases}
\tan^{2}\vartheta{}_{spec} & =\frac{\sin^{2}i+\sin^{2}e+2\sin i\sin e\cos\psi}{\left(\cos i+\cos e\right)^{2}}\\
\cos\zeta{}_{spec} & =\frac{1}{\sin i\tan\vartheta{}_{spec}}\left(\frac{\cos i+\cos e}{2}\left(1+\tan^{2}\vartheta{}_{spec}\right)-\cos i\right)\\
\sin\zeta{}_{spec} & =\frac{\sin e\sin\psi}{\left(\cos i+\cos e\right)\tan\vartheta}
\end{cases}\label{eq:tgv}
\end{equation}

\subsection{Expression of the specular reflectance for one pixel}
\label{sub:Expression-of-the}
We consider a pixel of area $A$ formed of $N$ facets of same area
$A_{f}$, orientated according the probability density $a\left(\vartheta,\zeta\right)$,
as detailed in section \ref{sub:Surface-rugosity}, with $N\gg1$.
The number of facets satisfying the specular reflection conditions
defined in section \ref{sub:Specular-conditions-for} will be $\iint_{\mathcal{H_{C}}}N\,\mbox{a}\left(\vartheta{}_{spec},\zeta{}_{spec}\right)\,\mbox{d\ensuremath{\left(\vartheta,\zeta\right)}}$,
where $\mathcal{H_{C}}$ is the set of values $\left(\vartheta,\zeta\right)$
satisfying \ref{eq:tgv} within the range of observation geometries.
Indeed, there is a range of different geometry of observation within
one instruments' pixel. Let $\chi_{c}$ be the emergence variations
within a pixel. The facet orientation satisfying the specular conditions
for this geometry is $\left(\vartheta_{spec},\zeta_{spec}\right)$
given by Eq. \ref{eq:tgv}. At the incidence $i$, there is a range
of orientations that satisfy specular condition within a pixel that
is centered at $\left(\vartheta_{spec},\zeta_{spec}\right)$ and has
the size $\delta\left(\vartheta,\zeta\right)$. $\delta\left(\vartheta,\zeta\right)$
is determined using a function 
\begin{equation*}
\mbox{g}_{i}:\,\begin{pmatrix}e\\
\psi
\end{pmatrix}\mapsto\begin{pmatrix}\mbox{g}_{1}\left(e,\psi\right)=\vartheta\\
\mbox{g}_{2}\left(e,\psi\right)=\zeta
\end{pmatrix} 
\end{equation*}
that transforms $\left(e,\psi\right)$ into $\left(\vartheta,\zeta\right)$,
for the incident angle $i$.
Not every facet that satisfy these conditions will send energy to
the captor. Indeed, the roughness of the surface introduces a shadowing
of the scene : some facets will not receive incident light, or will
not be visible by the captor, or both. A shadowing factor $\mbox{S}^{\prime}$
must be introduced at this point. Let $N_{spec}$ be the number of
facets that satisfy both the geometrical condition defined in section
\ref{sub:Specular-conditions-for} and the visibility condition :
\begin{equation}
N_{spec}=\iint_{\mathcal{H_{C}}}N\,\mbox{a}\left(\vartheta{}_{spec},\zeta{}_{spec}\right)\mbox{S}^{\prime}\left(i,e,\psi,\overline{\theta}\right)\,\mbox{d\ensuremath{\left(\vartheta,\zeta\right)}}
\end{equation}
with $\mbox{S}^{\prime}\left(i,e,\psi,\overline{\theta}\right)$ a
shadowing factor that depends on the geometry of observation and the
roughness of the surface \cite{Hapke1984}. Each one of these $N_{spec}$
facets receives an incident power $P_{i}$ and send back a reflected
power $P_{r}$ : 
\begin{equation}
P_{i}=FA_{f}\cos\left(\frac{\alpha^{\prime}}{2}\right)
\end{equation}
 
\begin{equation}
P_{r}=P_{i}\,\mbox{r}_{f}\left(\frac{\alpha^{\prime}}{2}\right)
\end{equation}
$F$ being the incident power flux (\textit{e.g.} : Solar flux) in
the radiation direction, $A_{f}\cos\left(\frac{\alpha^{\prime}}{2}\right)$
the projection of the facet in the plane orthogonal to the incident
radiation, and $\mbox{r}_{f}\left(\frac{\alpha^{\prime}}{2}\right)$
the Fresnel reflection coefficient in energy at the phase angle $\alpha^{\prime}$,
$\mbox{r}_{f}=R_{\text{\ensuremath{\perp}}}^{2}\left(\alpha\right)+R_{\parallel}^{2}\left(\alpha\right)$.
As $\alpha^{\prime}$ does not depend on the facet orientations, all
these specular reflections will result in a specular power $P_{spec}=N_{spec}P_{r}$,
thus :

\begin{multline}
P_{spec}\left(i,e,\psi,\overline{\theta}\right)=\iint_{\mathcal{H_{C}}}NFA_{f}\cos\left(\frac{\alpha^{\prime}}{2}\right)\mbox{r}_{f}\left(\frac{\alpha^{\prime}}{2}\right) \\ 
\mbox{a}\left(\vartheta{}_{spec},\zeta{}_{spec}\right)\mbox{S}^{\prime}\left(i,e,\psi,\overline{\theta}\right)\,\mbox{d\ensuremath{\left(\vartheta,\zeta\right)}}\label{eq:Pspec}
\end{multline}
The reflectance factor $R$ is the ratio between the bidirectional
reflectance $r$ of the surface and the one $r_{L}$ of a perfectly
lambertian surface , thus $R=\pi\frac{r}{\cos i}$. The bidirectional
reflectance $r$ is the ratio between the radiance $L$ of the surface
and the collimated incident power, perpendicularly to the incident
direction. Thus $r=\frac{L}{F}$, with $L=\frac{P}{\Omega A\cos e}$,
$A$ being the illuminated surface and $\Omega_{c}$ the solid angle
subtended by a pixel. Finally, $R_{spec}=\pi\frac{r_{spec}}{\cos i}=\pi\frac{L_{spec}}{F\cos i}=$$\pi\frac{P_{spec}}{\Omega_{c}AF\cos i\cos e}$,
thus : 
\begin{multline}
R_{spec}\left(i,e,\psi,\overline{\theta}\right)=\iint_{\mathcal{H_{C}}}\pi\frac{NA_{f}\cos\left(\frac{\alpha^{\prime}}{2}\right)}{\Omega_{c}A\cos i\cos e}\mbox{S}^{\prime}\left(i,e,\psi,\overline{\theta}\right) \\ 
\mbox{r}_{f}\left(\frac{\alpha^{\prime}}{2}\right)\mbox{a}\left(\vartheta{}_{spec},\zeta{}_{spec}\right)\,\mbox{d\ensuremath{\left(\vartheta,\zeta\right)}}\label{eq:dubidu}
\end{multline}
where $\Omega_{c}$ is the solid angle subtended by an instrument's
pixel. $A$ is the sum of the horizontal projection of all the facets
: $A=NA_{f}\left\langle \cos\vartheta\right\rangle $. The term $\left\langle \cos\vartheta\right\rangle $
is included in the shadowing function $\mbox{S}\left(i,e,\psi,\overline{\theta}\right)$
described by B. Hapke \cite{Hapke1984}. Thus we can simplify Eq. \ref{eq:dubidu}
as 
\begin{multline}
R_{spec}\left(i,e,\psi,\overline{\theta}\right)=\iint_{\mathcal{H_{C}}}\pi\frac{\cos\left(\frac{\alpha^{\prime}}{2}\right)}{\Omega_{C}\cos i\cos e}\mbox{S}\left(i,e,\psi,\overline{\theta}\right)\\ \mbox{r}_{f}\left(\frac{\alpha^{\prime}}{2}\right) \mbox{a}\left(\vartheta{}_{spec},\zeta{}_{spec}\right)\,\mbox{d\ensuremath{\left(\vartheta,\zeta\right)}}
\end{multline}
$\mbox{d\ensuremath{\left(\vartheta,\zeta\right)}}$ is derived from
the integration angles $e$ and $\psi$ that are the emergence and
azimuth angles. There is a bijection between $\Omega_{C}$ and $\mathcal{H_{C}}$
because \ref{eq:tgv} admits a unique solution for every $\left(e,\psi\right)$.
Considering that the incidence angle $i$ is a constant, we can rigorously
express $R_{spec}^{i}$ as : 
\begin{multline}
R_{spec}^{i}\left(i,e,\psi\right)=\frac{\pi}{\Omega_{c}}\iint_{\Omega_{c}}\frac{\cos\left(\frac{\alpha^{\prime}}{2}\right)}{\cos i\cos e}\,\mbox{S}\left(i,e,\psi,\overline{\theta}\right)\,\mbox{r}_{f}\left(\frac{\alpha^{\prime}}{2}\right) \\ \,\mbox{a}\left(\vartheta{}_{spec},\zeta{}_{spec}\right)\,\left|\det J_{\mbox{g}_{i}}\left(e,\psi\right)\right|\,\mbox{d\ensuremath{e}}\,\mbox{d}\ensuremath{\psi}\label{eq:Rspec_i_constant}
\end{multline}
$\left|\det J_{\mbox{g\ensuremath{_{e}}}}\left(e,\psi\right)\right|$
is the Jacobian of the function $\mbox{g}_{i}$. This expression~\ref{eq:Rspec_i_constant}
assumes that the incidence $i$ is a constant. In reality, the light
source is almost never completely collimated, but ranges inside a
solid angle (\textit{e.g. }: the Solar disk). Let $\Omega_{S}$ be
the solid angle of the source. Then the total specular contribution
within one pixel will be : 
\begin{equation}
R_{spec}\left(i,e,\psi\right)=\frac{1}{\Omega_{S}}\iint_{\Omega_{S}}R_{spec}^{i}\left(i,e,\psi\right)\,\left|\det J_{\mbox{g}_{e}}\left(i,\psi\right)\right|\sin i\,\mbox{d\ensuremath{i}}\,\mbox{d\ensuremath{\psi}}\label{eq:Rspec-pix}
\end{equation}
where $\left|\det J_{\mbox{g\ensuremath{_{e}}}}\left(i,\psi\right)\right|$
is the Jacobian of the function 
\begin{equation*}
\mbox{g}_{e}:\,\begin{pmatrix}i\\
\psi
\end{pmatrix}\mapsto\begin{pmatrix}\mbox{g}_{1}\left(i,\psi\right)=\vartheta\\
\mbox{g}_{2}\left(i,\psi\right)=\zeta
\end{pmatrix}
\end{equation*}
that transforms $\left(i,\psi\right)$ into $\left(\vartheta,\zeta\right)$,
for the given emergence angle $e$.

\section{Diffuse reflectance}

We consider a two layers model, with a slab overlaying a semi-infinite
granular substrate. The collimated radiation from the sun is transmitted
to the slab with an external reflection coefficient $S_{e}^{\prime}$
(the prime here represent the anisotropy). We suppose an isotropisation
at the second interface. The slab is modeled as a compact isotropic
and homogeneous matrix. It contains inclusions that are close to spherical
and not identical to the matrix. The inclusions are the main contributors
to the scattering of radiation in the layer. They are distributed
homogeneously in the matrix. The determination of the Fresnel coefficients
at the interface matrix/inclusion or inclusion/matrix is a key to
estimate the transmission and reflection factors of the layer. An
internal and external reflection coefficient $S_{ik}$ and $S_{ek}$
for each type of inclusion $k$ must be defined. 

In this part we describe the radiative transfer in the media. First
we will characterize the transmission of light into the slab. By energy
conservation this is equivalent to calculating the total reflected
power which, normalized by the incident energy, stands for the reflexion
coefficient (section~\ref{sub:reflection-coefficients-4.1}). Then
we will describe the scattering of light by the inclusion during the
transfer through the slab. This requires the calculation of the external
and internal reflection coefficients of these inclusions (section~\ref{sub:reflection-coefficients-4.2}).
Once the basic optical properties of the inclusions are known, we
can consider fluxes of energy within the whole slab that will be governed
by the radiative properties of the slab (section~\ref{sub:Radiative-properties-4.5}).
Solving this radiative transfer problem within the slab with an upper
and lower optical interfaces will give the overall reflection and
transmission factors of the slab (section~\ref{sub:reflection-and-transmission-4.6}).
Finally the radiative interaction of the two layers (substrate and
slab) are considered and solved by adding doubling leading to the
final result (section~\ref{sub:Bidirectionnal-reflectance-4.7}).

\subsection{Reflection coefficients for the slab}
\label{sub:reflection-coefficients-4.1}

\paragraph{Anisotropic case}

Let $S_{e}^{\prime}$ be the external reflection coefficient in a
collimated case (interface atmosphere/ice matrix). It corresponds
to the ratio between the incident power $P_{i}$ and the total reflected
power, in every direction $P_{r}^{tot}$. The total reflected power
can be estimated integrating the specular contributions for every
emerging direction, at the given incidence angle $i$. 
\begin{equation}
S_{e}^{\prime}=\frac{\iint_{2\pi}\,\mbox{d}P_{spec}}{AF\cos i}
\end{equation}
$\mbox{d}P_{spec}$ being the specular contribution in a given geometry.
Using Eq.  \ref{eq:Pspec}, the expression of $S_{e}^{\prime}$ becomes
\begin{equation}
S_{e}^{\prime}=\iint_{\mathcal{H}}\frac{\cos\left(\frac{\alpha^{\prime}}{2}\right)\,\mbox{r}_{f}\left(\frac{\alpha^{\prime}}{2}\right)\mbox{a}\left(\vartheta{}_{spec},\zeta{}_{spec}\right)\mbox{S}\left(i,e,\psi,\overline{\theta}\right)}{\cos i}\,\mbox{d\ensuremath{\left(\vartheta,\zeta\right)}}
\end{equation}
where $\mathcal{H}$ is the set of values taken by $\vartheta$ and
$\zeta$ throughout the integration. Exactly like in section \ref{sub:Expression-of-the},
$\mbox{d\ensuremath{\left(\vartheta,\zeta\right)}}$ is derived from
the integration angles $e$ and $\psi$ that are the emergence and
azimuth angles. There is now a bijection between $\mathcal{B}$ and
$\mathcal{H}$, $\mathcal{B}$ being the superior hemisphere that
is the domain of variation of $e$ and $\psi$. Considering that the
incidence angle $i$ is a constant, we can express $S_{e}^{\prime}$
as :

\begin{multline}
S_{e}^{\prime}=\int_{_{0}}^{^{\frac{\pi}{2}}}\int_{_{0}}^{^{2\pi}}\frac{\cos\left(\frac{\alpha^{\prime}}{2}\right)\,\mbox{r}_{f}\left(\frac{\alpha^{\prime}}{2}\right)\mbox{a}\left(\vartheta{}_{spec},\zeta{}_{spec}\right)\mbox{S}\left(i,e,\psi,\overline{\theta}\right)}{\cos i} \times \\ \left|\det J_{\mbox{g}_{i}}\left(e,\psi\right)\right|\,\mbox{d}e\,\mbox{d}\psi\label{eq:Sep}
\end{multline}
where $\left|\det J_{g}\left(e,\psi\right)\right|$ is the Jacobian
of the function 
\begin{equation*}
\mbox{g}_{i}:\,\begin{pmatrix}e\\
\psi
\end{pmatrix}\mapsto\begin{pmatrix}\mbox{g}_{1}\left(e,\psi\right)=\vartheta\\
\mbox{g}_{2}\left(e,\psi\right)=\zeta
\end{pmatrix} 
\end{equation*}
that transforms $\left(e,\psi\right)$ into $\left(\vartheta,\zeta\right)$,
for the incident angle $i$.

The internal reflection coefficient $S_{i}^{\prime}$ in a collimated
case at the interface ice matrix/atmosphere is not considered as we
suppose an isotropisation of the radiation at the second interface
(ice/granular regolith).

\paragraph{Isotropic case}

In the isotropic case, the internal reflection coefficient $S_{i}$
is obtained integrating the Fresnel equations at the surface for the
all geometries: 
\begin{equation}
S_{i}=\intop_{0}^{\frac{\pi}{2}}\left[R_{\text{\ensuremath{\perp}}}^{2}\left(\alpha\right)+R_{\parallel}^{2}\left(\alpha\right)\right]\cos\alpha\,\mbox{d\ensuremath{\alpha}}\label{eq:Si}
\end{equation}
where $R_{\perp}\left(\alpha\right)$ and $R_{\parallel}\left(\alpha\right)$
are Fresnel reflectivites for perpendicular and parallel polarization
with respects to the propagation plan, for an incidence angle $\alpha$
and will be detailed later.

The external reflection coefficient $S_{e}$ is estimated the same
way : 
\begin{equation}
S_{e}=\intop_{0}^{\frac{\pi}{2}}\left[R_{\text{\ensuremath{\perp}}}^{2}\left(\alpha\right)+R_{\parallel}^{2}\left(\alpha\right)\right]\cos\alpha\,\mbox{d\ensuremath{\alpha}}\label{eq:Se}
\end{equation}

\subsection{Reflection coefficients for the inclusions}
\label{sub:reflection-coefficients-4.2}
In the case of a spherical inclusion of the type $k$, the internal
reflection coefficient $S_{ik}$ is obtained in the usual fashion
integrating the Fresnel equations (see \cite{Hapkebook} \textit{Hapke},
2012, sect. 5.4.4, pp.78-95) 
\begin{equation}
S_{ik}=\intop_{0}^{\frac{\pi}{2}}\left[R_{\text{\ensuremath{\perp}}}^{2}\left(\alpha\right)+R_{\parallel}^{2}\left(\alpha\right)\right]\cos\alpha\sin\alpha\,\mbox{d\ensuremath{\alpha}}\label{eq:Si-1}
\end{equation}

For the estimation of the external reflection coefficient $S_{ek}$,
a differential absorption factor is taken into account. Indeed, as
we deal with inclusions in an absorbing matrix, the parallel rays
we consider in the integration touch the inclusion after different
optic paths. For a ray that touches the inclusion with an incidence
$\alpha$, the differential path length in the matrix is $\nu=\rho_{k}\cos\alpha$,
where $\rho_{k}$ is the radius of the spherical inclusion. Thus the
differential absorption factor is $e^{-a_{m}\rho_{k}\left(1-\cos\alpha\right)}$,
where $a_{m}$ is the absorption coefficient of the matrix. Writing
the matrix's optical index $n_{m}+i\, k_{m}$, the dispersion relation
gives $a_{m}=\frac{4\pi}{\lambda}k_{m}$. Finally, the external reflection
coefficient $S_{ek}$ at the interface matrix/inclusion is : 
\begin{equation}
S_{ek}=\intop_{0}^{\frac{\pi}{2}}\left[R_{\text{\ensuremath{\perp}}}^{2}\left(\alpha\right)+R_{\parallel}^{2}\left(\alpha\right)\right]\mbox{e}^{-a_{m}\rho\left(1-\cos\alpha\right)}\cos\alpha\sin\alpha\,\mbox{d\ensuremath{\alpha}}\label{eq:Se-1}
\end{equation}
This differential absorption effect is already taken into account
in the expression of $R_{\text{\ensuremath{\perp}}}\left(\alpha\right)$
and $R_{\parallel}\left(\alpha\right)$ in the case of the internal
reflection at the interface inclusion/matrix.

\subsection{Fresnel coefficients}
\label{sub:Fresnel-coefficients-4.3}
The Fresnel reflectivities for perpendicular and parallel polarization
with respects to the propagation plane, for an incidence angle $\alpha$,
$R_{\perp}\left(\alpha\right)$ and $R_{\parallel}\left(\alpha\right)$,
are derived from Snell's law (see \cite{Hapkebook} \textit{Hapke},
2012, sect. 4.3, pp.46-60) : 
\begin{equation}
R_{\perp}\left(\alpha\right)=\frac{\left(\cos\alpha-\mathcal{G}_{1}\right)^{2}+\mathcal{G}_{2}^{2}}{\left(\cos\alpha+\mathcal{G}_{1}\right)^{2}+\mathcal{G}_{2}^{2}}
\end{equation}
\begin{equation}
R_{\parallel}\left(\alpha\right)=\frac{\left[\left(n^{2}-k^{2}\right)\cos\alpha-\mathcal{G}_{1}\right]^{2}+\left[2nk\cos\alpha-\mathcal{G}_{2}\right]^{2}}{\left[\left(n^{2}-k^{2}\right)\cos\alpha+\mathcal{G}_{1}\right]^{2}+\left[2nk\cos\alpha+\mathcal{G}_{2}\right]^{2}}
\end{equation}
using $n=\frac{n_{1}n_{2}+k_{1}k_{2}}{n_{1}^{2}+k_{1}^{2}}$ and $k=\frac{n_{1}k_{2}-n_{2}k_{1}}{n_{1}^{2}+k_{1}^{2}}$,
with $n_{1}+ik_{1}$ and $n_{2}+ik_{2}$ the complex refractive indexes
of the media considered.
\begingroup\makeatletter\def\f@size{8.5}\check@mathfonts
\def\maketag@@@#1{\hbox{\m@th\large\normalfont#1}} 
\begin{equation}
\mathcal{G}_{1}^{2}=\frac{1}{2}\left[\left[n^{2}-k^{2}-\sin^{2}\alpha\right]+\left[\left(n^{2}-k^{2}-\sin^{2}\alpha\right)^{2}+4n^{2}k^{2}\right]^{\frac{1}{2}}\right]
\end{equation}
\begin{equation}
\mathcal{G}_{2}^{2}=\frac{1}{2}\left[-\left[n^{2}-k^{2}-\sin^{2}\alpha\right]+\left[\left(n^{2}-k^{2}-\sin^{2}\alpha\right)^{2}+4n^{2}k^{2}\right]^{\frac{1}{2}}\right]
\end{equation}
\endgroup

\subsection{Integration}
\label{sub:Integration-4.4}
S. Chandrasekhar showed (see \textit{Chandrasekhar}, 1960, sect. 22,
pp. 61-69 \cite{Chandrasekhar1960}) that many radiative transfer
integration can be approximated using the Gauss quadrature formulae.
If $f\left(\mu\right)$ is a polynomial of order $2m-1$, then 
\begin{equation}
\intop_{-1}^{1}f\left(\mu\right)\,\mbox{d}\mu=\sum_{j=1}^{m}c_{j}f\left(\mu_{j}\right)\label{eq:Gauss_quadrature}
\end{equation}
where $\mu_{1},...,\mu_{m}$ are the zeros of the Legendre polynomials
$P_{1},...,P_{m}$ of order $1,...,m$, and $c_{1},...,c_{j}$ are
the associated Christoffel numbers : 
\begin{equation}
c_{j}=\frac{1}{P_{m}^{\prime}\left(\mu_{j}\right)}\intop_{-1}^{1}\frac{P_{m}\left(\mu\right)}{\mu-\mu_{j}}\,\mbox{d}\mu
\end{equation}
Equation \ref{eq:Gauss_quadrature} is exact if $f\left(\mu\right)$
is a polynomial of order $2m-1$. When $f\left(\mu\right)$ is not
a polynomial, then the quadrature formulae gives an approximation
that converges to the exact value when $m\rightarrow\infty$. The
order $m$ of the approximation directly governs its quality. We estimate
analytically the internal and external reflection coefficients in
the isotropic case $S_{i}$ an $S_{e}$ using the roots of the 32$^{\mbox{th}}$
order Legendre's polynomial and the associated Christoffel's numbers
as detailed in \cite{davis1956abscissas}. We use a simple change
of variable to transform the integration interval from $[0,\frac{\pi}{2}]$
into $[-1,1]$. All the integrations are performed using the Gauss
quadrature formulae, except $S_{e}^{\prime}$ and $R_{spec}$. In
these cases, the integration being a double one, we cannot use the
Gauss quadrature. We chose after numerical tests an adaptive grid
and the rectangle method.

\subsection{Radiative properties of a slab containing inclusions}
\label{sub:Radiative-properties-4.5}
We suppose an homogeneous distribution of isotropic inclusions inside
the slab. The inclusions type is noted by $k$, with $N_{i}$ different
types, defined by different geometrical and optical properties.

\subsubsection{Proportions of inclusions }

We define the slab compactness $\gamma_{c}$ as the volume of the
ice matrix per unit of volume. We also define $\mathcal{N}$ the total
number of inclusions per unit of volume, and $\mathcal{N_{\textrm{k}}}$
the number of inclusions of the type $k$ per unit of volume. The
proportion of each type of inclusion is $P_{k}=\frac{\mathcal{N}_{k}}{\mathcal{N}}$.
Immediate geometrical calculations give : 
\begin{equation}
\mathcal{N}=\frac{3\left(\gamma_{c}-1\right)}{4\pi\sum_{k=1}^{N_{i}}P_{k}\rho_{k}}\label{eq:N}
\end{equation}

\subsubsection{Cross sections }

We suppose close to spherical inclusions. The scattering efficiency
for a sphere has been described by B. Hapke (\cite{Hapkebook} \textit{Hapke},
2012, sect. 5.6, pp.95-99, eq 5.52a) in his equivalent slab model.
For an inclusion of the type $k$ : 
\begin{equation}
Q_{sk}=S_{ek}+\left(1-S_{ek}\right)\frac{\left(1-S_{ik}\right)}{1-S_{ik}\Theta_{ik}}\Theta_{ik}
\end{equation}
where $S_{ik}$ and $S_{ek}$ are respectively the internal and external
reflection coefficients of an inclusion expressed in equations \ref{eq:Si-1}
and \ref{eq:Se-1}, and $\Theta_{ik}$ is the internal transmission
coefficient of and inclusion. In the two stream approximation, and
assuming the isotropy of the phase function of the internal scatterers
in an inclusion (\cite{Hapkebook} \textit{Hapke}, 2012, sect. 6.5,
pp.122-144, Eq.  6.26) the expression of $\Theta_{ik}$ can be reduced
simply to : 
\begin{equation}
\Theta_{ik}=\frac{r_{ik}+\exp\left(-\rho_{k}\sqrt{a_{ik}\left(a_{ik}+s_{ik}\right)}\right)}{1+r_{ik}\exp\left(-\rho_{k}\sqrt{a_{ik}\left(a_{ik}+s_{ik}\right)}\right)}
\end{equation}
 $a_{ik}$ being a type $k$ inclusion's absorption coefficient, $s_{ik}$
its scattering coefficient, and 
\begin{equation}
r_{ik}=\frac{1-\sqrt{\frac{a_{ik}}{a_{ik}+s_{ik}}}}{1+\sqrt{\frac{a_{ik}}{a_{ik}+s_{ik}}}}
\end{equation}
 The scattering cross section $\sigma_{sk}$ for one inclusion is
\begin{equation}
\sigma_{sk}=\sigma_{k}Q_{sk}
\end{equation}
where $\sigma_{k}$ is the geometrical cross section : $\sigma_{k}=\pi\rho_{k}^{2}$.
Let $\left\langle \sigma_{s}\right\rangle $ be the mean cross section
of the inclusions : 
\begin{equation}
\left\langle \sigma_{s}\right\rangle =\sum_{k=1}^{N_{i}}P_{k}\sigma_{sk}\label{eq:sigma}
\end{equation}
We do the approximation of geometric optics, so the extinction cross
section $\sigma_{ek}$ correspond to the geometrical cross section
$\sigma_{k}$.

\subsubsection{Single scattering albedo and optical thickness}
\label{sub:Single-scattering-albedo}
The single scattering albedo $\omega$ of an absorbing and scattering
object is defined as the ratio of the total amount of power scattered
to the total amount of power removed to the wave (absorbed or scattered).
We propose a simple statistical approach to express the single scattering
albedo of a unit of volume of slab containing inclusions. We use the
same method as \cite{Hapkebook} \textit{Hapke}, 2012, sect. 7.4,
pp.158-169, but we modify the medium description. After a travel of
length $\mbox{d\ensuremath{\nu}}$, the probability $p_{1}$ for a
photon to meet an inclusion and be scattered is: 
\begin{equation}
p_{1}=1-\exp\left(-\mathcal{N}\left\langle \sigma_{s}\right\rangle \frac{\ln\gamma_{c}}{\gamma_{c}-1}\mbox{d}\nu\right)
\end{equation}
The probability $p_{2}$ that this photon has not been absorbed by
the matrix before is:
\begin{equation}
p_{2}=\exp\left(-a_{m}\mbox{d}\nu\right)
\end{equation}
Thus the probability $p_{s}$ for a photon to be only scattered per
unit of length is: 
\begin{equation}
p_{s}=\frac{1}{\mbox{d}\nu}\exp\left(-a_{m}\mbox{d}\nu\right)\left[1-\exp\left(-\mathcal{N}\left\langle \sigma_{s}\right\rangle \frac{\ln\gamma_{c}}{\gamma_{c}-1}\mbox{d}\nu\right)\right]
\end{equation}
that becomes for an infinitesimal length $\mbox{d}\nu$:
\begin{equation}
p_{s}=\mathcal{N}\left\langle \sigma_{s}\right\rangle \frac{\ln\gamma_{c}}{\gamma_{c}-1}+\circ\left(1\right)\label{eq:s}
\end{equation}
equally, the probability $p_{3}$ for a photon to be absorbed or scattered
by an inclusion throughout $\mbox{d}\nu$ is: 
\begin{equation}
p_{3}=\exp\left(-a_{m}\mbox{d}\nu\right)\left[1-\exp\left(-\mathcal{N}\left\langle \sigma_{e}\right\rangle \frac{\ln\gamma_{c}}{\gamma_{c}-1}\mbox{d}\nu\right)\right]
\end{equation}
and the probability $p_{4}$ to be absorbed by the matrix during $\mbox{d}\nu$
is: 
\begin{equation}
p_{4}=1-\exp\left(-a_{m}\mbox{d}\nu\right)
\end{equation}
so the probability of extinction $p_{e}$ per unit of length is: 
\begin{equation}
p_{e}=\frac{1}{\mbox{d}\nu}\left[1-\exp\left(-\left(\mathcal{N}\left\langle \sigma_{e}\right\rangle \frac{\ln\gamma_{c}}{\gamma_{c}-1}+a_{m}\right)\mbox{d}\nu\right)\right]\label{eq:tau}
\end{equation}
when $\mbox{d}\nu$ is close to $0$, it becomes: 
\begin{equation}
p_{e}=\mathcal{N}\left\langle \sigma_{e}\right\rangle \frac{\ln\gamma_{c}}{\gamma_{c}-1}+a_{m}+\circ\left(1\right)
\end{equation}
 Finally we obtain the single scattering albedo of a slab containing
inclusions dividing $p_{s}$ by $p_{e}$: 
\begin{equation}
\omega=\frac{\mathcal{N}\left\langle \sigma_{s}\right\rangle }{\mathcal{N}\left\langle \sigma_{e}\right\rangle +\frac{\gamma_{c}-1}{\ln\gamma_{c}}a_{m}}\label{eq:omega}
\end{equation}
Equation \ref{eq:tau} gives the expression of the optical depth $\tau$
of a slab with inclusion : 
\begin{equation}
\tau=\left(\mathcal{N}\left\langle \sigma_{e}\right\rangle \frac{\ln\gamma_{c}}{\gamma_{c}-1}+a_{m}\right)\nu
\end{equation}

\subsection{Diffuse reflectance and transmission factors of the contaminated slab}
\label{sub:reflection-and-transmission-4.6}

\subsubsection{Diffuse reflectance of a slab under collimated illumination}

In this section, we suppose that the slab is under a collimated radiation.
As in section~\ref{sub:Surface-rugosity}, we suppose that the surface
is constituted of $N$ unresolved facets that have a slope distribution
given by the probability density function $\mbox{a}\left(\vartheta,\zeta\right)$.
Each facet will receive an illumination at an incidence $i_{f}$ depending
on its orientation. We consider in that case that the first transit
in the slab is collimated and will transmit rays of light into the
slab at different inclinations. Our goal at this point is to determine
the mean transmission path $\overline{D^{\prime}}$ trough a slab
of a given roughness $\bar{\theta}$ and a given thickness $D$. For
a facet with the orientation defined by $\left(\vartheta,\zeta\right)$
using Snell-Descartes law gives $D^{\prime}=\mbox{fact\ensuremath{\left(\vartheta,\zeta\right)}}D$,
with:
\begingroup\makeatletter\def\f@size{8}\check@mathfonts
\def\maketag@@@#1{\hbox{\m@th\large\normalfont#1}}
\begin{equation}
\mbox{fact}\left(\vartheta,\zeta\right)=\frac{1}{\left|-\frac{1}{n_{m}}\cos i+\cos\vartheta\left(\frac{1}{n_{m}}\cos i_{f}-\sqrt{1-\frac{1}{n_{m}^{2}}\left(1-\cos^{2}i_{f}\right)}\right)\right|}\label{eq:fact}
\end{equation}
\endgroup
$i_{f}$ being the incidence angle on the facet (\textit{i.e. }the
angle between the facet's normal and the incident radiation). Using
basic trigonometric relations gives \cite{Hapkebook}: 
\begin{equation}
\cos i_{f}=\sin i\sin\vartheta\cos\zeta+\cos\vartheta\cos i
\end{equation}
We consider that only the first transit in the slab is anisotropic.
The internal absorption factor for the first anisotropic transit $\Theta^{\prime}$
will depend on the mean length $L^{\prime}$ of this transit, with
$\begin{aligned}\overline{D^{\prime}}= & \mbox{\ensuremath{\overline{\mbox{fact}}}}D\end{aligned}
$, and 
\begingroup\makeatletter\def\f@size{8}\check@mathfonts
\def\maketag@@@#1{\hbox{\m@th\large\normalfont#1}}
\begin{equation}
\begin{aligned}\overline{\mbox{fact}} & =\end{aligned}
\intop_{0}^{2\pi}\intop_{0}^{\frac{\pi}{2}}\mbox{fact}\left(\vartheta,\zeta\right)\,\mbox{a}\left(\vartheta,\zeta\right)\,\mbox{d\ensuremath{\vartheta}}\,\mbox{d\ensuremath{\zeta}}
\end{equation}
thus
\begin{multline}
\begin{aligned}\overline{\mbox{fact}} & =\frac{1}{\pi^{2}\tan\bar{\theta}}\end{aligned}\times \\
\intop_{0}^{2\pi}\intop_{0}^{\frac{\pi}{2}}\frac{e^{-\frac{\tan^{2}\vartheta}{\pi\tan^{2}\bar{\theta}}}\sec^{2}\vartheta\sin\vartheta}{\left|-\frac{1}{n_{m}}\cos i+\cos\vartheta\left(\frac{1}{n_{m}}\cos i_{f}-\sqrt{1-\frac{1}{n_{m}^{2}}\left(1-\cos^{2}i_{f}\right)}\right)\right|}\,\mbox{d\ensuremath{\vartheta}}\,\mbox{d\ensuremath{\zeta}}
\end{multline}
\endgroup
and 
\begin{equation}
\Theta^{\prime}=\frac{r_{m}+\exp\left(-D^{\prime}\sqrt{a_{m}\left(a_{m}+p_{s}\right)}\right)}{1+r_{m}\exp\left(-D^{\prime}\sqrt{a_{m}\left(a_{m}+p_{s}\right)}\right)}
\end{equation}
where $p_{s}$ is given by Eq.  \ref{eq:s}, $r_{m}=\frac{1-\sqrt{1-\omega}}{1+\sqrt{1-\omega}}$,
and $\omega$ is given by Eq.  \ref{eq:omega}. 

The internal absorption factor for an isotropic transit is (\cite{Hapkebook},
\textit{Hapke}, 2012 Eq.  6.26) 
\begin{equation}
\Theta=\frac{r_{m}+\exp\left(-2D\sqrt{a_{m}\left(a_{m}+s\right)}\right)}{1+r_{m}\exp\left(-2D\sqrt{a_{m}\left(a_{m}+s\right)}\right)}
\end{equation}
 Every following transit is considered isotropic. As illustrated on
figure~\ref{fig:Illustration-of-radiative}, we can express the reflectance
of the slab under a collimated radiation $R_{0}^{\prime\prime}$ as
\begin{equation}
R_{0}^{\prime\prime}=S_{e}^{\prime}+\left(1-S_{e}^{\prime}\right)\Theta^{\prime}S_{i}\Theta\left(1-S_{i}\right)\left[1+\sum_{n=1}^{\infty}\left(\Theta S_{i}\right)^{2}\right]
\end{equation}
$S_{i}$ being the internal reflection coefficient of the slab. The
term $S_{e}^{\prime}$ represents the integration over the sky of
the specular reflectance, and the other represents the diffuse reflectance.
Thus we can express the diffuse reflectance of the slab as 
\begin{equation}
R_{0}^{\prime}=\frac{\left(1-S_{e}^{\prime}\right)\Theta^{\prime}S_{i}\Theta\left(1-S_{i}\right)}{1-\left(\Theta S_{i}\right)^{2}}
\end{equation}
The diffuse transmission of the slab under a collimated radiation
$T_{0}^{\prime}$ is obtained the same way : 
\begin{equation}
T_{0}^{\prime}=\frac{\Theta^{\prime}\left(1-S_{e}^{\prime}\right)\left(1-S_{i}\right)}{1-\left(\Theta S_{i}\right)^{2}}
\end{equation}

\subsubsection{Diffuse reflectance of a slab under isotropic illumination }

In this section we suppose that the slab is under an isotropic radiation.
Indeed, at the lower interface, it is illuminated isotropically from
below by the substratum. $R_{0}$ and $T_{0}$ have their usual expressions
in this case : 
\begin{equation}
R_{0}=S_{e}+\frac{\left(1-S_{e}\right)S_{i}\Theta^{2}\left(1-S_{i}\right)}{1-\left(\Theta S_{i}\right)^{2}}
\end{equation}
\begin{equation}
T_{0}=\frac{\Theta\left(1-S_{e}\right)\left(1-S_{i}\right)}{1-\left(\Theta S_{i}\right)^{2}}
\end{equation}

\subsection{Bidirectional reflectance of a contaminated slab overlaying a semi-infinite granular media}
\label{sub:Bidirectionnal-reflectance-4.7}

In realistic conditions, a slab will receive a collimated radiation
from the solar disk, and a diffuse radiation from the granular medium
underneath. There is a coupling between the two layers, illustrated
on Figure~\ref{fig:Illustration-of-radiative}. Using adding doubling
formulas \cite{Doute1998}, we can express the total diffuse reflectance
of the slab over a granular substrate as :
\begin{align}
R_{Diff} & =R_{0}^{\prime}+T_{0}^{\prime}T_{0}r_{s}\sum_{n=0}^{\infty}\left(R_{0}r_{s}\right)^{n}\nonumber \\
 & =R_{0}^{\prime}+\frac{T_{0}^{\prime}T_{0}r_{s}}{1-R_{0}r_{s}}\label{eq:Rdiff}
\end{align}
where $r_{s}=\frac{1-\sqrt{1-\omega_{s}}}{1+\sqrt{1-\omega_{s}}}$
is the lambertian reflectance of the substrate \cite{Doute1998}.
$\omega_{s}$ is the single scattering albedo of the granular substrate.
The last step is to simulate the diffuse contribution for one measurement.
The total reflectance (BRDF) of the surface measured by the instrument
is the sum of the specular and diffuse contributions : 
\begin{equation}
R_{tot}=R_{spec}+R_{Diff}
\end{equation}
where $R_{spec}$ is determined by Eq.  \ref{eq:Rspec-pix} and $R_{Diff}$
by Eq.  \ref{eq:Rdiff}.

\section{Discussion}

\subsection{Energy conservation}
\label{sub:Energy-conservation}

\paragraph*{At the first interface}

\begin{figure}[htbp]
\centering
\fbox{\includegraphics[width=\linewidth]{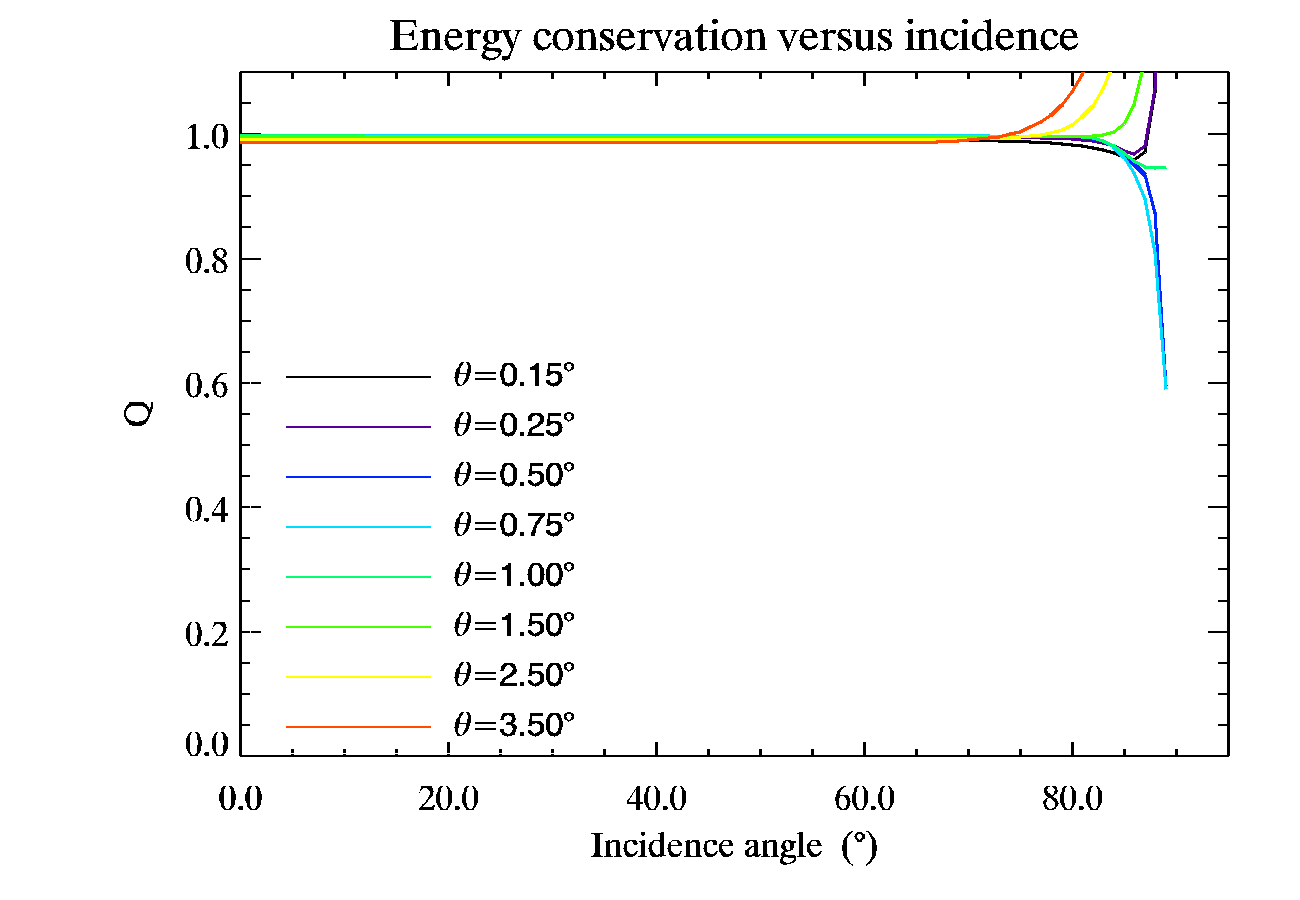}}
\caption{$Q$ as a function on the incidence angle $i$, when forcing the value of the Fresnel's reflection coefficient $\mbox{r}_{f}$ in Eq. \ref{eq:Sep} to one. A value of one means that
the conservation of energy is respected. This figure shows that in the cases of a roughness parameter ranging from $\bar{\theta}=0.15\text{\textdegree}$ to $\bar{\theta}=3.5\text{\textdegree}$, the energy is fairly well conserved for incidences below $85\text{\textdegree}$, and a roughness parameter below $\bar{\theta}=2.5\text{\textdegree}$. Thus, this model will not be applicable to very high incidences values, and roughness over $\bar{\theta}=2.5\text{\textdegree}$}
\label{fig:Cons_ener_Sep}
\end{figure}

We checked the conservation of the energy at different points in the
model. We first checked it at the first interface, as it contains
a complex numerical integration. To test the conservation of the energy
at the first interface, we force the value of the Fresnel's reflection
coefficient $\mbox{r}_{f}$ in Eq. \ref{eq:Sep} to one. Thus, all
the energy is supposed to be sent back, and we have to obtain $Q=1$
to have the energy conserved, where 
\begingroup\makeatletter\def\f@size{8.7}\check@mathfonts
\def\maketag@@@#1{\hbox{\m@th\large\normalfont#1}}
\begin{equation}
Q=\int_{_{0}}^{^{\frac{\pi}{2}}}\int_{_{0}}^{^{2\pi}}\frac{\cos\left(\frac{\alpha^{\prime}}{2}\right)\mbox{a}\left(\vartheta{}_{s},\zeta{}_{s}\right)\mbox{S}\left(i,e,\psi,\overline{\theta}\right)}{\cos i}\left|\det J_{\mbox{g}_{i}}\left(e,\psi\right)\right|\,\mbox{d}e\,\mbox{d}\psi
\end{equation}
\endgroup
that is Eq. \ref{eq:Sep} where the Fresnel reflection coefficient
is put to one. Figure~\ref{fig:Cons_ener_Sep} shows the value $Q$
as a function of the incidence angle. Different roughness parameters
$\bar{\theta}$ were tested, ranging from $\bar{\theta}=0.01\text{\textdegree}$
to $\bar{\theta}=45\text{\textdegree}$. Only values ranging from
$\bar{\theta}=0.15\text{\textdegree}$ to $\bar{\theta}=3.5\text{\textdegree}$
are displayed on Figure~\ref{fig:Cons_ener_Sep}. This test illustrates
the dependance of the validity of the model on both the incidence
angle and the roughness parameter.

\paragraph*{For the complete model}

\begin{figure}[htbp]
\centering
\fbox{\includegraphics[width=\linewidth]{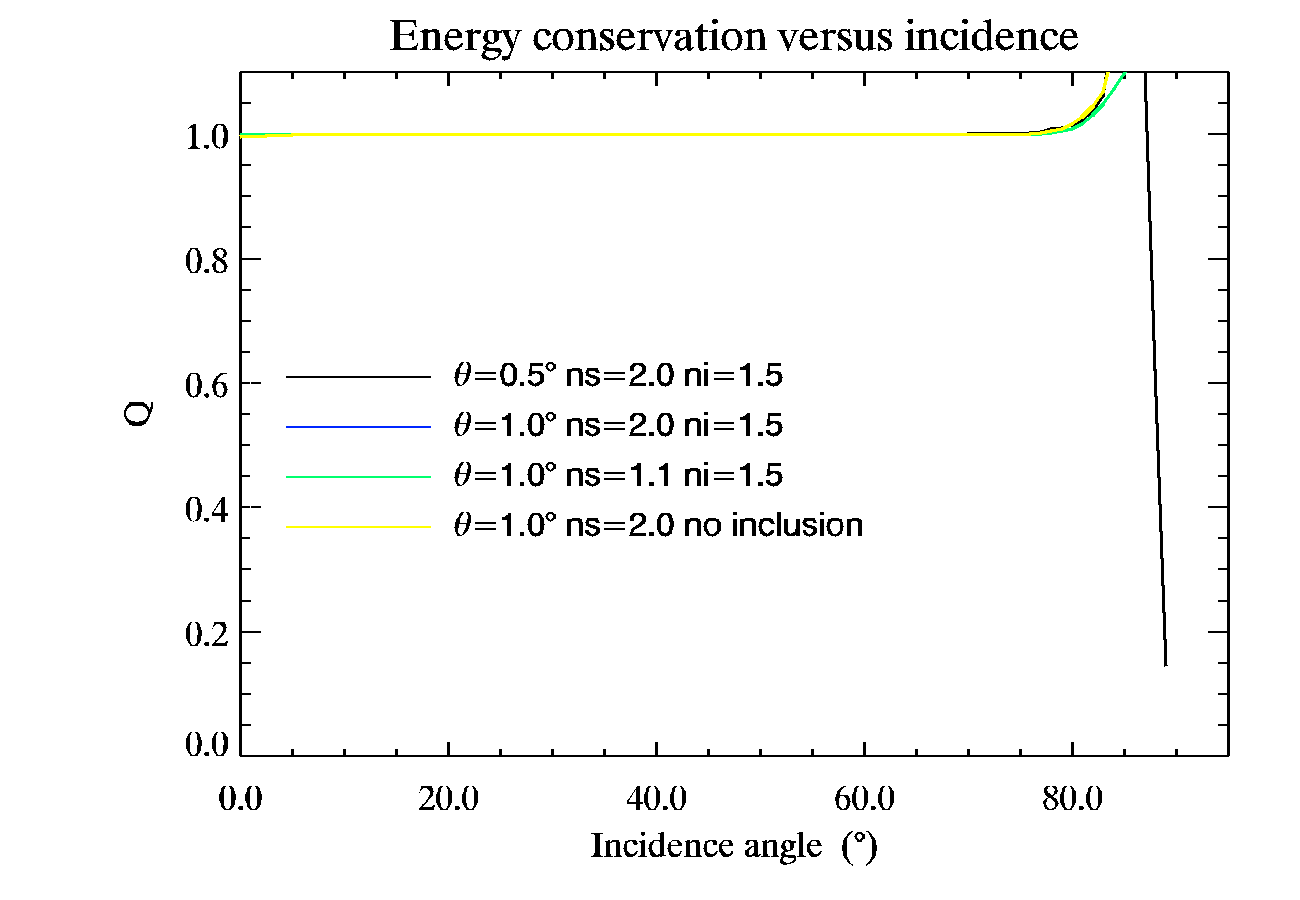}}
\caption{$Q$ as a function on the incidence angle $i$, where $E_{r}$ is the energy sent back from the surface, integrated over the hemisphere, and $E_{i}$ the incident energy.}
\label{fig:cons_ener_tot}
\end{figure}

To test the conservation of the energy for the whole model, we first
have to set the complex value of the optical constant of the slab
and the granular substrate to $0$, to make the surface non absorbent.
Then we integrate the energy sent back toward the sky. This energy
must equal the incoming energy. To test this practically in the model,
we set the sensor's angular aperture to a value that is equal to the
integration step. Figure~\ref{fig:cons_ener_tot} show the value
of $Q=\frac{2}{\pi}\intop_{0}^{\pi}\intop_{0}^{\frac{\pi}{2}}R_{tot}\cos e\sin e\,\mbox{d}e\,\mbox{d\ensuremath{\psi}}$.
Practically, the energy is conserved if this integral equals $1$.
Indeed, the energy conservation gives 
\begin{equation}
\intop_{sky}\frac{LA\cos e}{FA\cos i}\,\mbox{d}\Omega=1
\end{equation}
where $L$ is the surface radiance ($\mbox{W}.\mbox{m}^{-2}.\mbox{sr}^{-1}$)
$A$ is the surface of a pixel , $F$ is the incident flux in the
incident direction ($\mbox{W}.\mbox{m}^{-2}$) and $i$ and $e$ are
the incidence and emergence angles. The relation $R=\pi\frac{L}{F\cos i}$
between the reflectance factor and the radiance brings 
\begin{equation}
\frac{1}{\pi}\intop_{0}^{2\pi}\intop_{0}^{\frac{\pi}{2}}R_{tot}\cos e\sin e\,\mbox{d}e\,\mbox{d\ensuremath{\psi}}=1
\end{equation}
The symmetry of the model in azimuth leads to the quantity $Q$ displayed
in Figure~\ref{fig:cons_ener_tot}. This figure shows that it is
mostly the roughness parameter $\bar{\theta}$ and the incidence angle
$i$ that control the validity of the model.

\begin{figure}[htbp]
\centering
\fbox{\includegraphics[width=\linewidth]{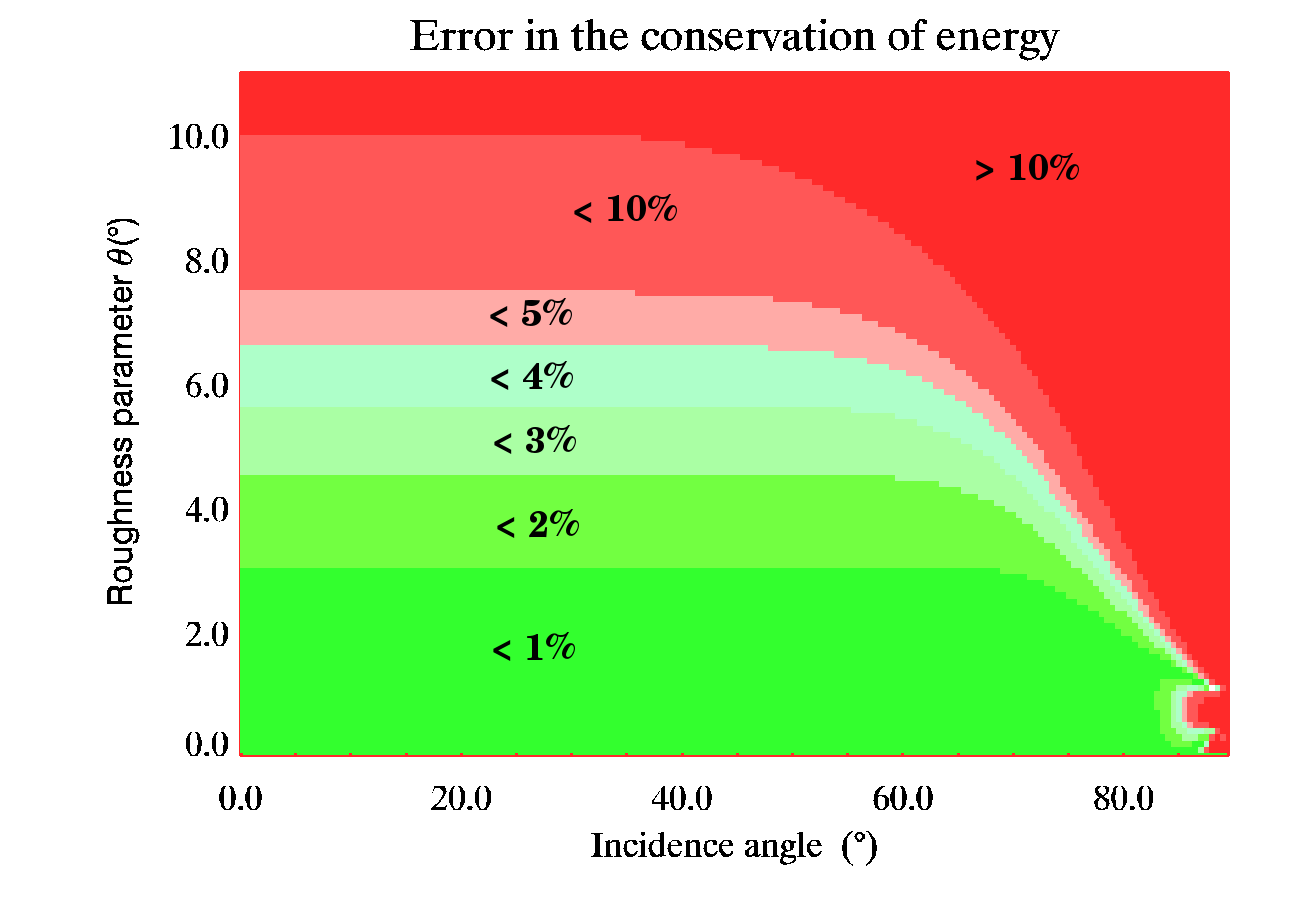}}
\caption{Error in the energy conservation as a function of $\bar{\theta}$ and $i$. }
\label{fig:Error-in-cons} 
\end{figure}

Figure~\ref{fig:Error-in-cons} shows the error in the energy conservation
in percent, as a function of the roughness parameter $\bar{\theta}$
and the incidence $i$. This gives the range of validity of the model
according to a given tolerance. Roughness parameters larger than $\bar{\theta}=11\text{\textdegree}$, always exhibiting error larger than $10\,\mbox{\%}$, are not represented.
For small slab real optical index (\textit{i.e. }close to one), these
errors decrease.

\subsubsection{Slope distribution}

\begin{figure}[htbp]
\centering
\fbox{\includegraphics[width=\linewidth]{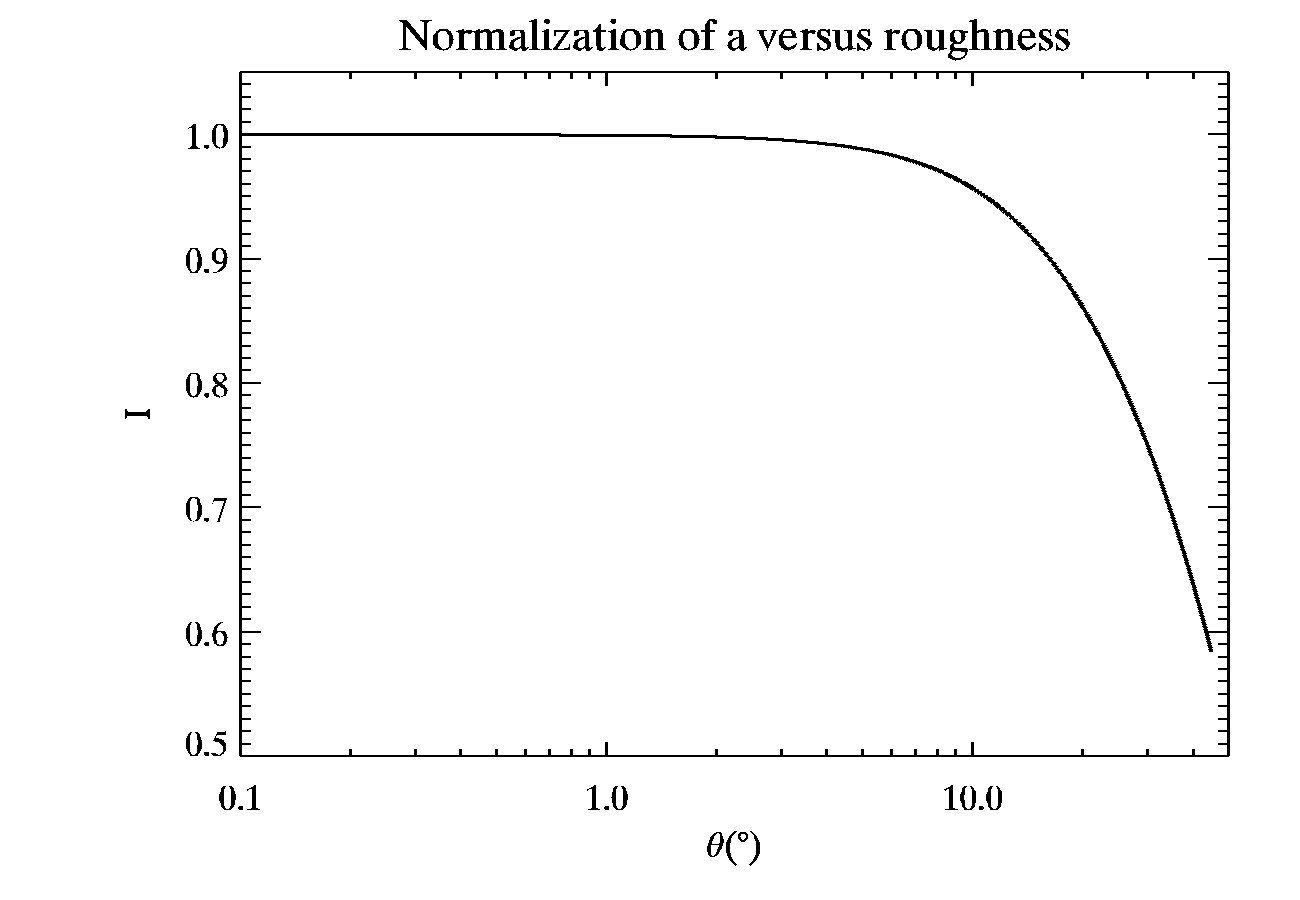}}
\caption{$I$ versus $\bar{\theta}$. For values of $\bar{\theta}$ larger than $\bar{\theta}=2\text{\textdegree}$, the probability density function for the slopes begins to drop. At $\bar{\theta}=10\text{\textdegree}$, $I=0.957$.}
\label{fig:Proba_slope}
\end{figure}

As mentioned in section~\ref{sub:Surface-rugosity}, this model is
limited to the case of small $\bar{\theta}$. Figure~\ref{fig:Error-in-cons}
gives a quantification of that limitation. This is mostly due to the
fact that the probability density function $\mbox{a}\left(\vartheta,\zeta\right)$
that defines the repartition of slopes only makes sense if $\iint_{\left(\vartheta,\zeta\right)}\mbox{a}\left(\vartheta,\zeta\right)\mbox{d\ensuremath{\vartheta}}\,\mbox{d\ensuremath{\zeta}}=1$,
which means that 
\begin{equation}
\frac{1}{2\pi}\int_{_{0}}^{^{\frac{\pi}{2}}}\mbox{a}\left(\vartheta\right)\mbox{d\ensuremath{\vartheta}}=1
\end{equation}
or that the value of $I=\int_{_{0}}^{^{\frac{\pi}{2}}}\frac{2}{\pi\tan^{2}\overline{\theta}}\exp\left(-\frac{\tan^{2}\vartheta}{\pi\tan^{2}\overline{\theta}}\right)\sec^{2}\vartheta\sin\vartheta\,\mbox{d\ensuremath{\vartheta}}$
must be equal to $1$. Figure~\ref{fig:Proba_slope} shows the value
of \textit{$I$ }versus the roughness parameter $\bar{\theta}$. The
function $\mbox{a}\left(\vartheta,\zeta\right)$ only makes sense
as a probability function if $I=1$. For values of roughness larger
than $\bar{\theta}=2\text{\textdegree}$, $I$ begins to fall to values
below one. In a further development, we could extend the applicability
of the model by defining a new probability density function $\mbox{a}_{Norm}\left(\vartheta,\zeta\right)$,
that would be the normalization of the function $\mbox{a}\left(\vartheta,\zeta\right)$
: $\mbox{a}_{Norm}\left(\vartheta,\zeta\right)=\frac{1}{I\left(\bar{\theta}\right)}\mbox{a}\left(\bar{\theta,}\vartheta,\zeta\right)$. 

\subsection{Behavior of the model}

\subsubsection{Specular reflection}

\begin{figure}[htbp]
\centering
\fbox{(a)\includegraphics[width=\linewidth]{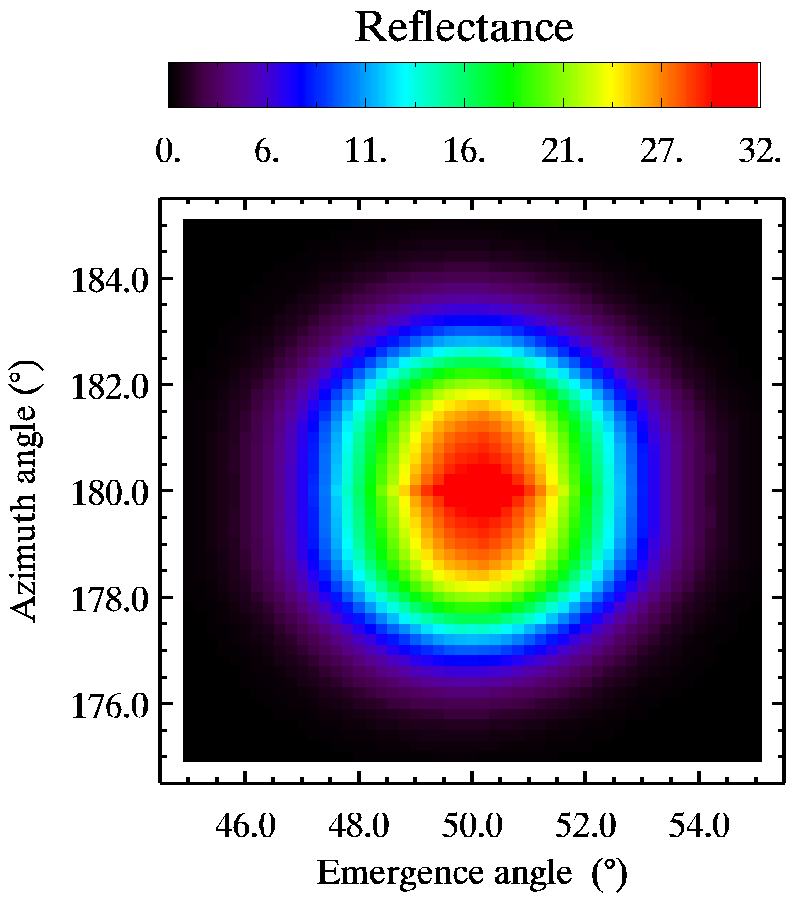}}
\fbox{(b)\includegraphics[width=\linewidth]{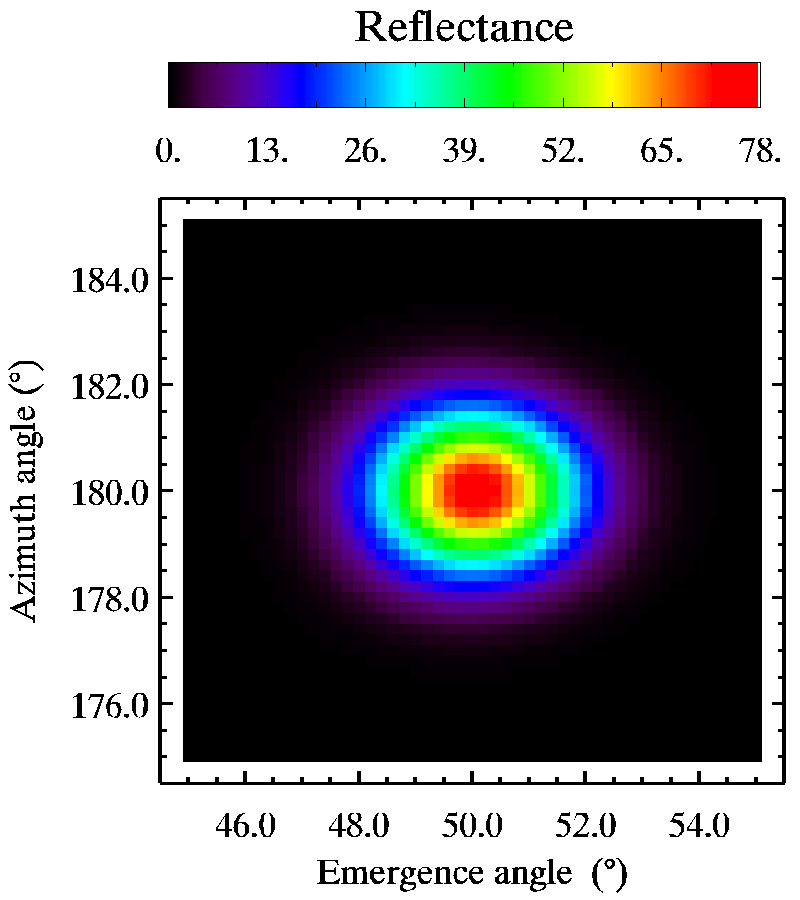}}
\caption{Zoom on the specular spot for a water ice slab
at $1\,\mbox{\textmu m}$ with a roughness parameter $\bar{\theta}=0.5\text{\textdegree}$,
(a) illuminated at an incidence angle $i=50\text{\textdegree}$ with
a light source that has an aperture of $0.4\text{\textdegree}$,
and observed with a detector that has an aperture of $4.2\text{\textdegree}$,
such as in the conditions of a laboratory measurement with the instrument
described in \cite{Brissaud2004}, and (b) illuminated at an incidence
angle $i=50\text{\textdegree}$ with a light source that has an aperture
of $0.2\text{\textdegree}$, and observed with a detector that
has an aperture of $6.92.10^{-2}\text{\textdegree}$, such as in the
conditions of a measure with the high resolution spectro imaging imaging
instrument OMEGA orbiting Mars described in \cite{Bibring_et_al}.}
\label{fig:Specular}
\end{figure}

This model is built to simulate observations. Thus, the specular spot
characteristics will depend not only on the illumination divergence
and the geometry, but also on the observation device. This model is
designed to be adaptable to both conditions. Figure~\ref{fig:Specular}
shows a zoom on the specular spot for a water ice slab at $1\,\mbox{\textmu m}$
with a roughness parameter $\bar{\theta}=0.5\text{\textdegree}$,
illuminated at an incidence angle $i=50\text{\textdegree}$ illuminated
with different light sources, and observed with two distinct detectors.
In the first case (in Figure \ref{fig:Specular}a), the surface
is illuminated with a light source that has an aperture of $0.4\text{\textdegree}$
and observed with a sensor that has an aperture of $4.2\text{\textdegree}$.
It represents the conditions of a laboratory measurement with the
instrument described in \cite{Brissaud2004}. In the second case (in
Figure \ref{fig:Specular}b), the surface is illuminated with
a light source that has an aperture of $0.2\text{\textdegree}$ and
observed with a sensor that has an aperture of $6.92.10^{-2}\text{\textdegree}$.
It represents the conditions of a measure with the OMEGA imaging spectrometer
instrument orbiting the planet Mars \cite{Bibring_et_al}. Both cases
represent actual measurement situations. As shown on Figure~\ref{fig:Specular},
both the amplitude and the shape of the specular spot depend on the
characteristics of the illumination. 

\subsubsection{Influence of the parameters }

\begin{figure}[htbp]
\centering
\fbox{\includegraphics[width=\linewidth]{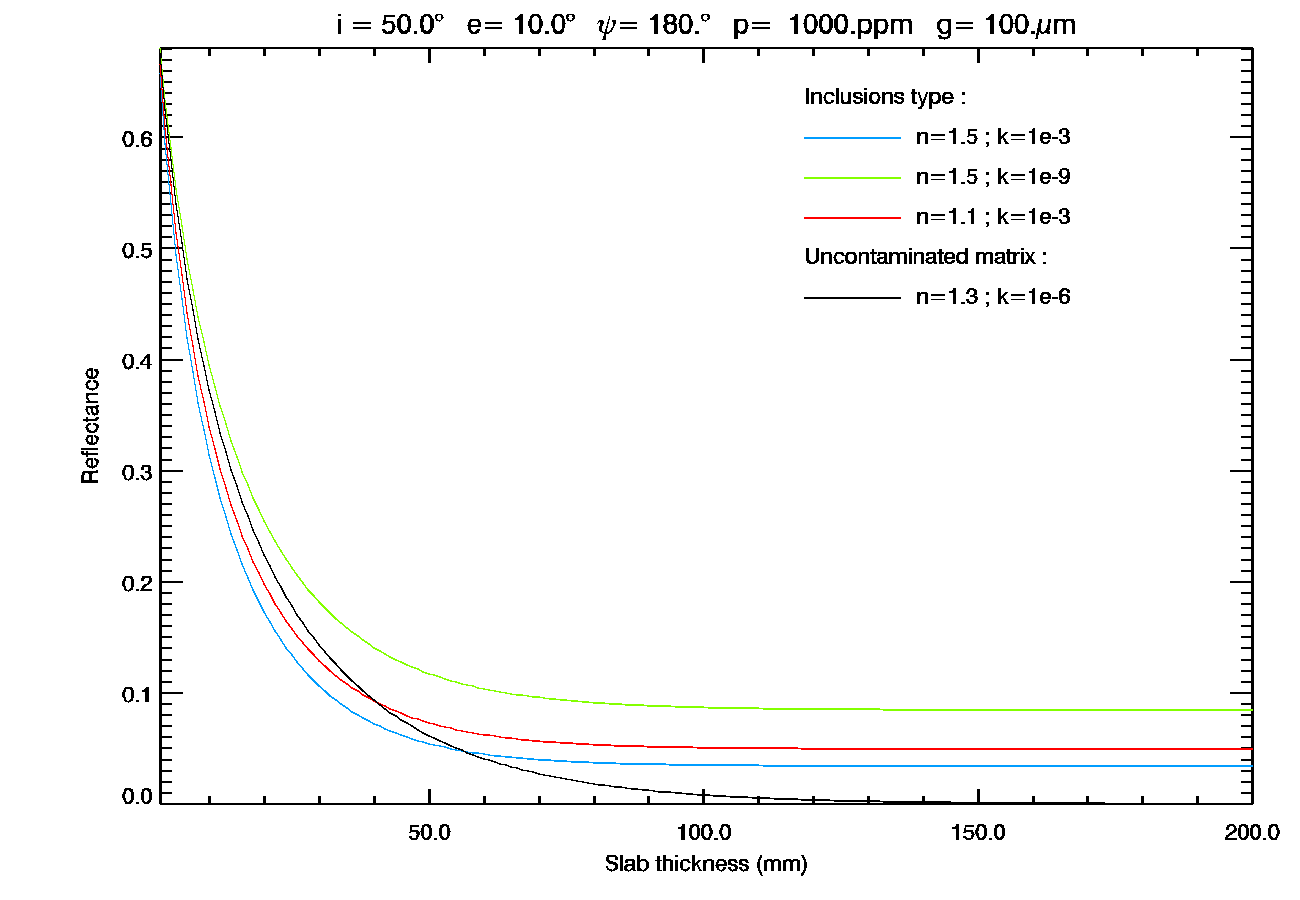}}
\caption{Reflectance factor of a slab of water ice containing various types of inclusions, at a wavelength $\lambda=1\,\text{\textmu m}$, as a function of the thickness of the slab layer, other parameters fixed. Black curve is the uncontaminated reference, and colored curves represent different optical indexes of inclusions.}
\label{fig:Influence-of-h}
\end{figure}

\begin{figure}[htbp]
\centering
\fbox{\includegraphics[width=\linewidth]{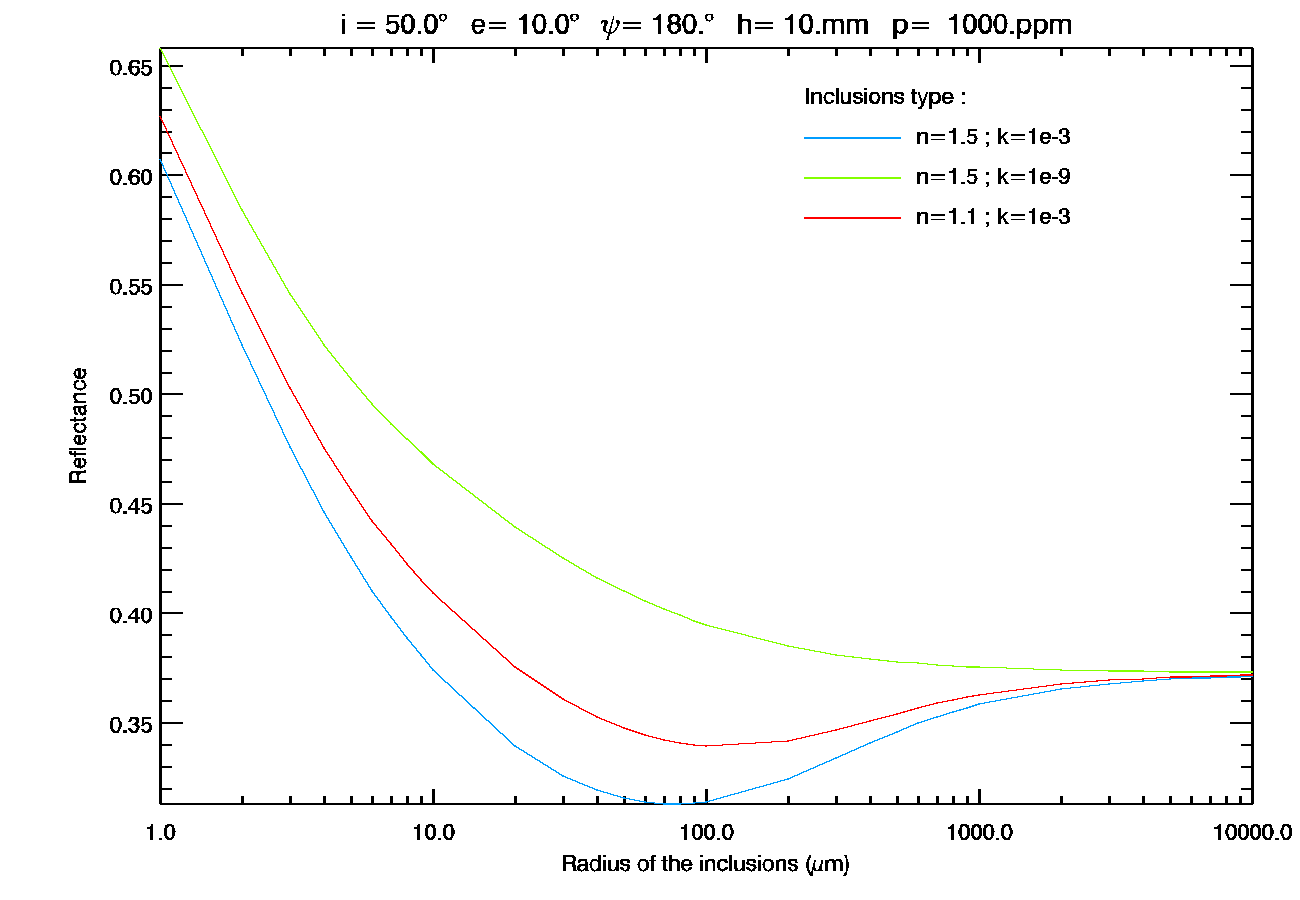}}
\caption{Reflectance factor of a slab of water ice containing various types of inclusions, at a wavelength $\lambda=1\,\text{\textmu m}$, as a function of the grain size of the inclusions, other parameters fixed. Colored curves represent different optical indexes of inclusions. When the absorption coefficient is higher in the inclusions than in the matrix (blue and red curves), there is a competition between scattering and absorption. The dominant effect depend on the grain size. The reflectance factor of an uncontaminated at this geometry is approximately $R=0.372$.}
\label{fig:Influence-of-g}
\end{figure}

\begin{figure}[htbp]
\centering
\fbox{(a)\includegraphics[width=\linewidth]{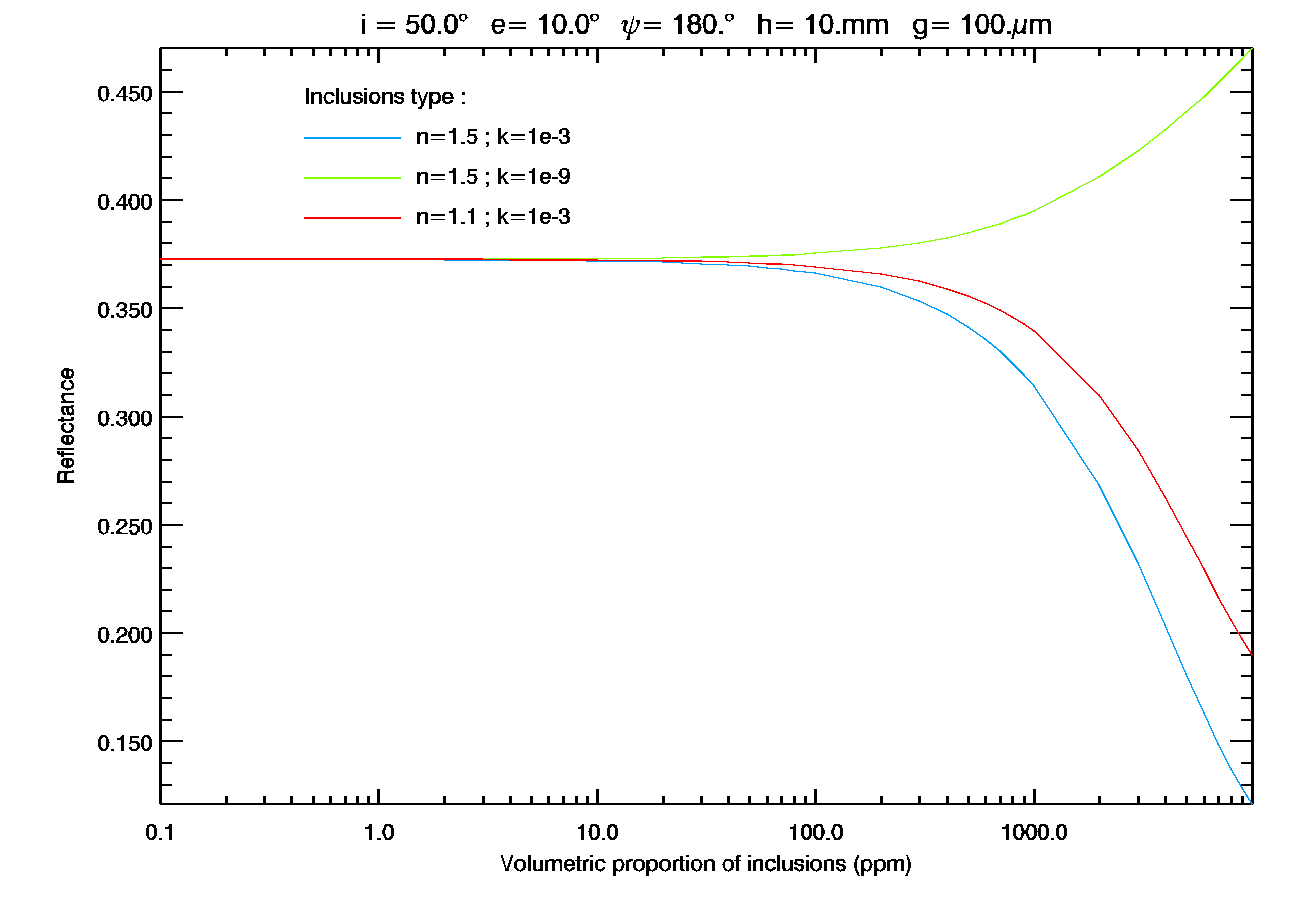}}
\fbox{(b)\includegraphics[width=\linewidth]{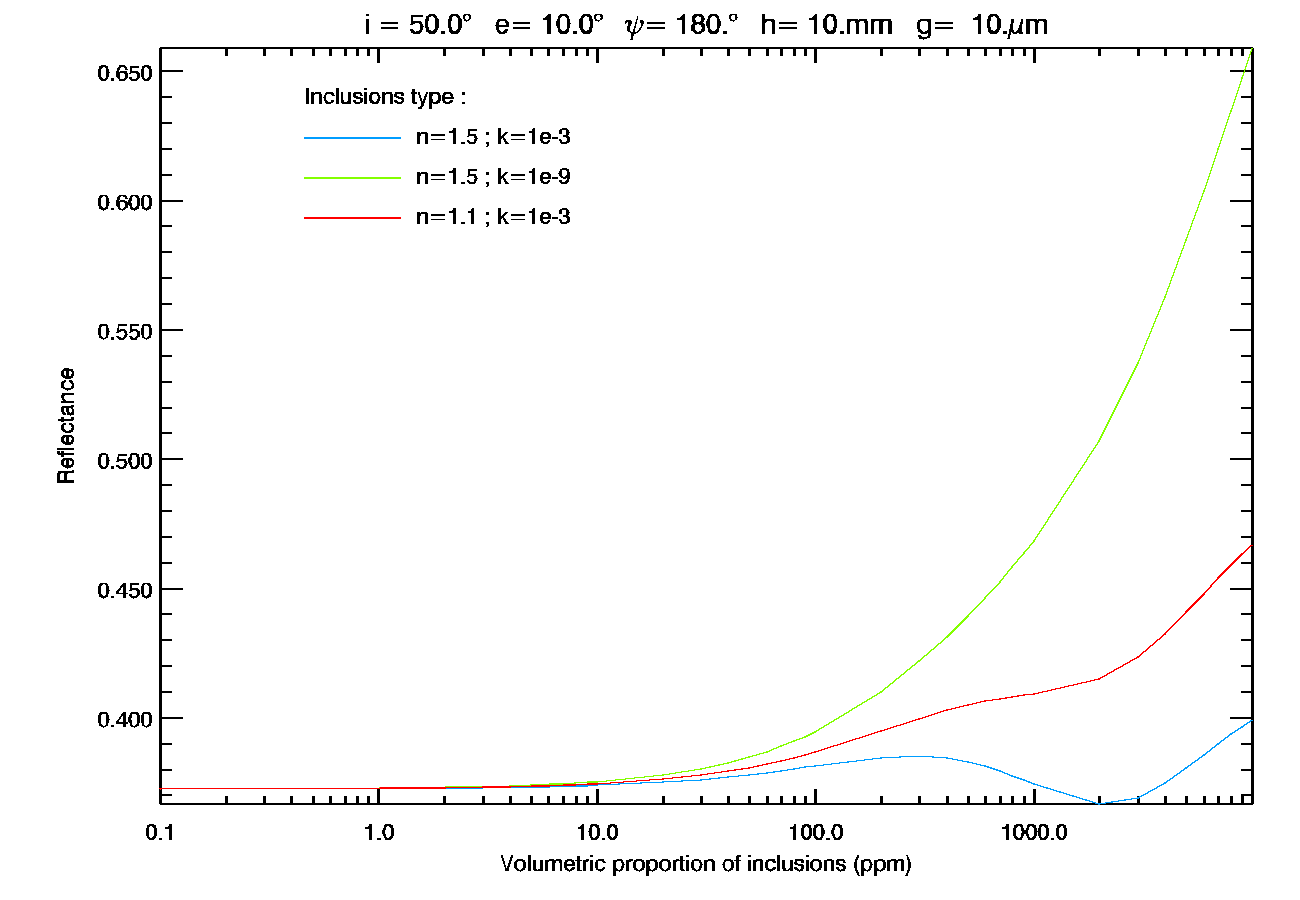}}
\caption{Reflectance factor of a slab of water ice containing various types of inclusions, at a wavelength $\lambda=1\,\text{\textmu m}$, as a function of the volumetric proportion of inclusions, other parameters fixed. Colored curves represent different optical indexes of inclusions. (a) absorption is the dominant effect for blue and red curves. (b) scattering is the dominant effect for red curve, and scattering and absorption are of the same order of magnitude (blue curve). The reflectance factor of an uncontaminated at this geometry is approximately $R=0.372$.}
\label{fig:Influence-of-c}
\end{figure}

To give a feeling on how the model behaves according to the different
parameters, we chose a set of parameters, and plotted the dependence
of the reflectance to the variation of one parameter around this first
set. We chose arbitrary optical constants for the matrix and inclusions.
We chose for the matrix $n=1.3$ and $k=1.10^{-3}$, that are approximately
the values for water ice at $270\,\mbox{K}$, and at the $1\,\text{\textmu m}$
wavelength. We selected as our standard set of parameter a $10\,\mbox{mm}$
thick slab layer containing $1000\,\mbox{ppmv}$ of $100\,\mbox{\textmu m}$
wide inclusions, overlaying a semi infinite granular layer of the
same nature that the matrix. We tested the behavior of the model for
various types of inclusions. We describe two types of behavior. The
first type is when the absorption coefficient of the inclusions is
smaller than the one of the matrix. This includes the particular case
of a matrix contaminated with bubbles. It is illustrated in the figure~\ref{fig:Influence-of-h},
\ref{fig:Influence-of-c} and \ref{fig:Influence-of-g} with the green
curves. In this case, the real part $n$ of the optical index of the
inclusions has very little influence. The second case is when the
absorption in the inclusion is bigger than in the matrix (blue and
red curves). In this case, the real optical index of the inclusions
plays an important role. 

\begin{figure}[htbp]
\centering
\fbox{\includegraphics[width=\linewidth]{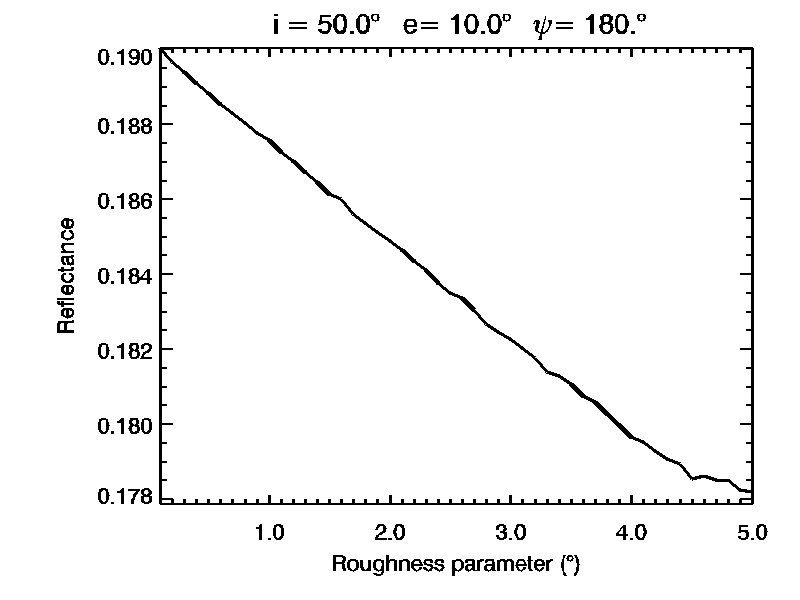}}
\caption{Reflectance factor of a $20\,\text{mm}$ thick slab of water ice containing $1000\,\text{ppm}$ of $100\,\text{\textmu m}$ wide inclusions as a function of the roughness parameter $\bar{\theta}$, other parameters fixed. The optical indexes of the inclusions are $n=1.1$ and $k=1.10^{-3}$. The reflectance decrease as the roughness increase. This is due to the fact that a bigger roughness means more facets in specular conditions and thus less energy inserted into the system. The roughness parameter have a smaller impact on the diffuse reflectance than the other parameters of the model. }
\label{fig:Influence-of-theta}
\end{figure}

Figure~\ref{fig:Influence-of-h} shows the dependence of the reflectance
on the thickness of the slab layer in different cases. In the case
of an uncontaminated slab, the reflectance approaches $0$ when the
thickness increases. On the contrary, the reflectance of a slab containing
inclusions will saturate at a value depending on the properties of
the impurities. In the first case of a low absorption in the inclusions
(green curve) the reflectance is higher than the reflectance of an
uncontaminated layer whatever thickness the slab has. On the other
case of high absorption in the inclusions, when the slab layer is
thin, then the reflectance is lower than one of an uncontaminated
slab, but as thickness increases, the value saturates, due to the
scattering of light by the inclusions. This value gives an idea of
the penetration depth of the light into a contaminated slab layer
(\textit{i.e. }the depth from which the layer becomes optically thick).
Figure~\ref{fig:Influence-of-g} show the dependence of the reflectance
on the radius of inclusions in the slab layer. This illustrates the
scattering properties of the inclusions. The reflectance factor of
an uncontaminated at this geometry is approximately $R=0.372$. In
this figure, the volumetric proportion of inclusions in the matrix
is held constant, so as the grain size increases, the number of inclusions
per unit of volume decreases. Thus the scattering power decreases
as well as a function of the grain size. In the case of inclusions
with higher absorption than the matrix, at some point, the grain size
reaches a value where the absorption in the inclusions becomes more
efficient than the scattering effect, and the reflectance falls below
the reference value of the uncontaminated slab (around $10\,\mbox{\textmu m}$
for the blue curve and $20\,\mbox{\textmu m}$ for the red one). Then,
the decreasing probability of encountering an inclusion when grain
sizes become too high make the reflectance approach the value of the
uncontaminated slab. Indeed, when the grain size of the inclusion
equals the thickness of the layer (at the extreme right of the plot),
the probability of encountering one, knowing the the volumetric proportion
is $1000\,\mbox{ppmv}$ becomes very low and the influence of the
inclusions become negligible. 

Figure~\ref{fig:Influence-of-c} shows the evolution of the reflectance
in different cases where scattering or absorption dominates. On Figure~\ref{fig:Influence-of-c}a,
the absorption is the dominant effect. In the case of inclusions with
a higher absorption coefficient than the matrix, this makes the reflectance
drop when the proportion of inclusions increases (blue and red curves).
For the green curve, both absorption and scattering contribute in
increasing the reflectance of the slab, thus it increases with the
proportion of inclusions. Figure~\ref{fig:Influence-of-c}b shows more
complexity. It is the same as Figure~\ref{fig:Influence-of-c}a except
that the grain size of the inclusions is $10\,\text{\textmu m}$ instead
of $100\,\text{\textmu m}$. The green curve still represents the
case of a lower absorption in the inclusion. Absorption and diffusion
both tend to increase the reflectance, thus it increases with the
proportion of inclusions. The red curve represents a case of higher
absorption in the inclusions, when the scattering contribution is
dominant. In this case, as diffusion limits the penetration of light
into the layer, the reflectance increases with the proportion of impurities.
The blue curve represents the limit case where diffusion and absorption
contributions are of the same order of magnitude, leading to strong
non-linear behavior. 

Figure~\ref{fig:Influence-of-theta} shows the dependence  of the reflectance of a $20\,\text{mm}$ thick slab of water ice containing $1000\,\text{ppm}$ of $100\,\text{\textmu m}$ wide inclusions on the roughness parameter $\bar{\theta}$. The optical indexes of the inclusions are $n=1.1$ and $k=1.10^{-9}$. The diffuse reflectance of a slab decreases as the roughness increases. A bigger roughness means a bigger diversity in the slope distribution at the surface. This leads to an increased number of facets satisfying the specular reflection conditions defined in section~\ref{sub:Specular-conditions-for}. Finally, less energy is inserted into the surface, and the diffuse reflectance is smaller. The dependance of the diffuse reflectance on the roughness if smaller compared to the dependance on the other parameters. This can be attributed to the relatively small range of values of $\bar{\theta}$. On the contrary the roughness parameter $\bar{\theta}$ has a strong influence on the specular contribution. 

\section{Conclusions}

We developed a radiative transfer model to simulate the bidirectional
reflectance of a rough slab with inclusions. Typical calculation time
is $1.10^{-2}\,\mbox{s}$ per spectrum, considering $10000$ wavelengths.
Nevertheless, it can vary greatly depending on the sets of parameters
desired. 

Most of the constituting elements of this model have already been
numerically \cite{Doute1998} or experimentally validated \cite{Hapkebook}
but we adapted them to build a new model. In this study we tested
numerically the conservation of energy and characterized the domain
of validity of the model. We conducted sensibility studies in the
case of a matrix containing only one type of inclusion. This shows
the complexity and non linearity of the model with respect to its
parameters. The sensibility study in the case of several types of
inclusions were not conducted because of the great number of parameters.
The experimental validation will be conducted in a following paper. 

This model is designed to analyze massive hyperspectral data in the
planetary science domain. In our favorite application, it calculates
the radiative transfer in a contaminated ice slab overlaying an optically
thick granular medium. The contamination in the slab can be of any
type : ice, minerals or even bubbles. The matrix can be constituted
as well of any ice. Thus our model can be applied on Earth with water
ice, but also on Mars polar region covered with CO$_{2}$ ice \cite{Leighton1966},
on icy bodies, such as Jupiter's moon Europa (water ice), Neptune's
moon Triton or dwarf planet Pluto (N$_{2}$ ice). Other applications
in biology or industry are possible, as soon as the optical constants
of each material are known. 

We considered in all the calculations every wavelength independently.
Thus, a spectrum in any spectral range can be built by computing every
wavelength contribution at very high spectral resolution. The final
objective is the comparison of the simulation to actual data, for
analysis purposes. This makes this approach suitable for any spectroscopic
measurement of slabs (made of ice or other material), overlaying optically
thick material (granular or other material), from laboratory to spatial
probe measurement. For the planetary science case, these results will
be down-sampled at the instrument's wavelength resolution, using its
PSFs. 

One major hypothesis in this work is that we suppose an isotropic
behavior of the inclusions. In the future, we plan to add the particles
phase function to improve this point. We also plan to normalize the 
probability distribution function describing the roughness of the
surface, to extend the applicability of the model.

\bigskip
\noindent

% Bibliography
%\bibliography{../../../../Biblio}

\appendix

\section*{Notations}

\begin{tabular}{l p{.8\linewidth}} 
$\alpha^{\prime}$ & phase angle (\textdegree{})  \\  
$\gamma_{c}$ & compactness of the matrix : volume of matrix per unit of volume  \\  
$\zeta$ & slope azimuth angle (\textdegree{})  \\  
$\zeta_{spec}$ & slope azimuth angle of a facet in specular conditions (\textdegree{})  \\  
$\bar{\theta}$ & roughness parameter (\textdegree{})  \\  
$\vartheta$ & slope angle (\textdegree{})  \\  
$\vartheta_{spec}$ & slope angle of a facet in specular conditions(\textdegree{})  \\  
$\Theta_{ik}$ & transmission factor of a type k  inclusion  \\  $\Theta$ & transmission factor of the slab containing inclusion under isotropic illumination  \\  
$\Theta^{\prime}$ & transmission factor of the slab containing inclusion under collimated illumination  \\  
$\nu$ & optical path ($\mbox{m}$)  \\  
$\rho_{k}$ & radius of a type k  inclusion ($\mbox{m}$)  \\  
$\sigma_{k}$ & geometrical cross section for type k  inclusions ($\mbox{m}^{2}$)  \\  
$\sigma_{ek}$ & extinction cross section for type k  inclusions ($\mbox{m}^{2}$)  \\  
$\left\langle \sigma_{e}\right\rangle $ & mean extinction cross section ($\mbox{m}^{2}$)  \\  
$\sigma_{sk}$ & scattering cross section for type k  inclusions ($\mbox{m}^{2}$)  \\  
$\left\langle \sigma_{e}\right\rangle $ & mean scattering cross section ($\mbox{m}^{2}$)  \\  
$\tau$ & optical depth of the matrix containing inclusions  \\  
$\psi$ & azimuth angle (\textdegree{})  \\  
$\omega$	 & single scattering albedo of the matrix containing inclusions  \\ 
 $\omega_{s}$ & single scattering albedo of the granular substrate  \\ 
$\Omega_{C}$	& Solid angle subtended by the sensor ($\mbox{sr}$)  \\  
$\Omega_{S}$	& Solid angle subtended by the source ($\mbox{sr}$)  \\ 
$a_{m}$ & absorption coefficient of the matrix  \\  
$a_{ik}$ & absorption coefficient a type k  inclusion  \\  
$\mbox{a}\left(\vartheta,\zeta\right)$	& probability of occurrence for the slope$\left(\vartheta,\zeta\right)$  \\  $A$	& Surface of a pixel ($\ensuremath{m^{2}}$)  \\  
$A_{f}$ & Surface of a facet ($\ensuremath{m^{2}}$)  \\  
$D$ & thickness of the slab layer ($\mbox{m}$)  \\  $D^{\prime}$ & apparent length of the first transit through the slab layer for one ray ($\mbox{m}$)  \\  
$\overline{D^{\prime}}$ & mean apparent length of the first transit through the slab layer for one ray ($\mbox{m}$)  \\  
$e$ & emergence angle (\textdegree{})  \\  
$F$ & incident power flux in the radiation's direction ($\mbox{W}.\mbox{m}^{-2}$) \\  
$i$ & incidence angle (\textdegree{})  \\  
$k_{m}$ & Imaginary part of the optical index of the matrix  \\  
\end{tabular}
\begin{tabular}{l p{.8\linewidth}} 
$k_{ik}$ & Imaginary part of the optical index of a type k  inclusion  \\ 
$L$ & radiance ($\mbox{W}.\mbox{m}^{-2}.\mbox{sr}^{-1}$)  \\  
$n_{m}$ & Real part of the optical index of the matrix  \\  
$n_{ik}$ & Real part of the optical index of a type k  inclusion  \\  
$N$ & number of facets within a pixel : $N\gg1$  \\  
$N_{spec}$ & number of facets within a pixel satisfying specular conditions  \\  
$\mathcal{N}$ & total density of inclusions inside the matrix  \\  
$\mathcal{N}_{k}$ & density of inclusions of type k  inside the matrix  \\  
$p$ & probability or probability per unit of length  \\  
$P$ & power ($\mbox{W}$)  \\  
$Q_{sk}$ & scattering efficiency for type k  inclusions  \\  
$r$ & bidirectional reflectance ($\mbox{sr}^{-1}$)  \\  
$r_{ik}$ & diffusive reflectance for a type k  inclusion   \\  
$r_{m}$ & diffusive reflectance of the matrix  \\  
$r_{s}$ & diffusive reflectance of the granular substrate  \\  
$\mbox{r}_{f}$ & Fresnel reflection coefficient $\mbox{r}_{f}=R_{\perp}^{2}+R_{\parallel}^{2}$  \\  
$R_{\perp}$ & Fresnel reflectivites for perpendicular polarization  \\  
$R_{\parallel}$ & Fresnel reflectivites for perpendicular and parallel polarization  \\  
$R_{0}$ & reflection factor of the slab containing inclusion under isotropic illumination  \\  
$R_{0}^{\prime\prime}$ & reflection factor of the slab containing inclusion under collimated illumination  \\  
$R_{0}^{\prime}$ & $R_{0}^{\prime\prime}-S_{e}^{\prime}$  \\  
$R_{Diff}$	 & diffuse reflectance factor of a slab containing inclusion over a granular substrate  \\  
$R_{spec}$ & specular reflectance factor of a slab containing inclusion over a granular substrate  \\  
$R_{tot}$ & reflectance factor of a slab containing inclusion over a granular substrate \\  
$\mbox{S}$ & shadowing function  \\  
$S_{e}^{\prime}$ & external reflection coefficient of the slab under collimated illumination \\  
$S_{e}$ & external reflection coefficient of the slab under isotropic illumination  \\  
$S_{i}$ & internal reflection coefficient of the slab under isotropic illumination  \\  
$S_{ek}$ & external reflection coefficient of a type k  inclusion  \\  
$S_{ik}$ & internal reflection coefficient of a type k  inclusion  \\  
$T_{0}$ & transmission factor of the slab containing inclusion under isotropic illumination  \\  
$T_{0}^{\prime}$ & transmission factor of the slab containing inclusion under collimated illumination \\  
\end{tabular}

\end{document}